\input harvmac
\input epsf

\def\calh{{\cal H}}
\def\G{{\cal G}}

\def\H{{\cal H}}
\def\e{\epsilon}

\def\Z{{\bf Z}_4}

\def\dzm{{\partial}}
\def\dzp{{\bar \partial}}

\def\a {{\alpha}}

\def\b {{\beta}}
\def\g {{\gamma}}
\def\d {{\delta}}
\def\s {{\sigma}}

\def\ah {{\widehat\alpha}}
\def\mh {{\widehat\mu}}
\def\bh {{\widehat\beta}}
\def\gh {{\widehat\gamma}}

\def\ad {{\dot\alpha}}

\def\bd {{\dot\beta}}
\def\gd {{\dot\gamma}}

\def\adh {{{\widehat{\dot\alpha}}}}
\def\mdh {{\widehat{\dot\mu}}}
\def\bdh {{\widehat{\dot\beta}}}
\def\gdh {{\widehat{\dot\gamma}}}

\def\t {{\theta}}
\def\tb {{\bar\theta}}
\def\ta {{\theta^\alpha}}
\def\tba {{\bar\theta^\ad}}
\def\th {{\widehat\theta}}
\def\tbh {{\widehat{\bar\theta}}}
\def\p {{\partial}}

\Title{ \vbox{\baselineskip12pt
              \hbox{IFT-P-060/99}
              \hbox{HUTP-99/A044}
              \hbox{CTP-MIT-2878}
              \hbox{hep-th/9907200} } }
{\vbox{
\centerline{Superstring Theory on $AdS_2 \times S^2$} 
\vskip .1in
\centerline{as a Coset Supermanifold}}}
\centerline{N. Berkovits$^a$, M. Bershadsky$^b$,
T. Hauer$^c$, S. Zhukov$^b$ and B. Zwiebach$^c$}
\vskip .1in
\centerline{$^{(a)}$\it Instituto de F\'{\i}sica Te\'orica,
Universidade Estadual Paulista}
\centerline{\it Rua Pamplona 145, 01405-900,
S\~ao Paulo, SP, Brasil}
\vskip .1in
\centerline{$^{(b)}$\it Lyman Laboratory of Physics, Harvard
University}
\centerline{\it Cambridge, MA 02138, USA}
\vskip .1in
\centerline{$^{(c)}$\it Center for Theoretical Physics,
Massachusetts Institute of Technology}
\centerline{\it Cambridge, MA 02138, USA}
\vskip .1in

\centerline{\bf Abstract}

We quantize the superstring on the $AdS_2\times S^2$ background with
Ramond-Ramond flux using a $PSU(1,1|2)/U(1)\times U(1)$ sigma model with a
WZ term.  One-loop conformal invariance of the model is guaranteed by a
general mechanism which holds for coset spaces $G/H$ where $G$ is
Ricci-flat and $H$ is the invariant locus of a Z${}_4$ automorphism of
$G$. This mechanism gives conformal theories for the $PSU(1,1|2)\times
PSU(2|2)/ SU(2)\times SU(2)$ and $PSU(2,2|4)/ SO(4,1)\times SO(5)$ coset
spaces, suggesting our results might be useful for quantizing the
superstring on $AdS_3\times S^3$ and $AdS_5\times S^5$ backgrounds.

\vskip .1in
\Date{July 1999}

\newsec{Introduction}

Two dimensional sigma models with target space {\it supermanifolds}
naturally appear in the quantization of superstring theory with
Ramond-Ramond (RR) backgrounds.  These sigma models possess manifest
target
space supersymmetry and have no worldsheet spinors. The fermionic
fields
are worldsheet scalars as in the usual Green-Schwarz (GS) formalism.
It
was recently shown that a sigma model on the supergroup manifold
$PSU(1,1|2)$ can be used for quantizing superstring theory on the
$AdS_3\times S^3$ background with RR flux
\nref\bvw{N. Berkovits, C. Vafa and  E. Witten,
{\it Conformal Field Theory
of AdS Background with Ramond-Ramond Flux}, hep-th/9902098,
JHEP 9903 (1999).}  
\nref\bzv{M. Bershadsky, S. Zhukov and A. Vaintrob, 
{\it $PSL(n|n)$ Sigma Model as a Conformal Field Theory},
hep-th/9902180}
\refs{\bvw,\bzv}.
In this paper, we show that the sigma model defined on the coset
supermanifold $PSU(1,1|2)/U(1)\times U(1)$ can be used to quantize
superstring theory on the $AdS_2\times S^2$ background with RR flux.
Furthermore, we show that this sigma model is one-loop conformal
invariant
and we expect that it is conformal to all orders in perturbation
theory.
Our methods can also be used to produce one-loop conformal sigma models
for
$AdS_3\times S^3$ and $AdS_5\times S^5$ backgrounds.  In the last case,
incorporation of the model into a consistent superstring theory remains
to
be done.

\nref\ts{R. Metsaev and A. Tseytlin, {\it Type IIB superstring action
in
$AdS_5 \times S^5$ background}, hep-th/9805028, Nucl.Phys. {\bf B533}
(1998) 109.}
\nref\tfol{ R. Kallosh, J. Rahmfeld and A. Rajaraman, {\it Near
Horizon Superspace}, hep-th/9805217, JHEP 9809 (1998) 002\semi R.
Kallosh,
J. Rahmfeld, {\it The GS String Action on $AdS_5 \times S^5$},
hep-th/9808038, Phys.Lett. {\bf B443} (1998) 143\semi I. Pesando, {\it
A k
Gauge Fixed Type IIB Superstring Action On ${AdS}_5\times S^5$}, JHEP
11
(1998) 002, hep-th/9808020.}  
\nref\tftwo{J.-G. Zhou,
{\it Super 0-brane and GS Superstring Actions on $AdS_2\times S^2$},
hep-th/9906013.}
\nref\tfthree{
I. Pesando, {\it The GS Type IIB Superstring Action On $AdS_3\times S^3
\times  T^4$}, hep-th/9809145\semi
J. Rahmfeld and A. Rajaraman, {\it The GS String Action On $
AdS_3\times
S^3$ with Ramond-Ramond Charge}, hep-th/9809164 \semi J. Park and
S.-J. Rey, {\it Green-Schwarz Superstring on $AdS_3\times {S}^3$},
hep-th/9812062.}

There are two conventional approaches for describing the superstring,
neither of which has been useful for quantizing the superstring in
backgrounds with RR flux. One such approach is the covariant GS
formalism
where spacetime-supersymmetry is manifest and worldsheet supersymmetry
is
absent.  The covariant GS superstring action can be defined classically
in
any background which satisfies the supergravity equations of
motion. When
the background fields satisfy these equations, the GS action is
classically
invariant under $\kappa$-symmetry which is necessary for removing
unphysical fermionic degrees of freedom.  Recently, this classical
action
was explicitly constructed for the case of the $AdS_5\times S^5$
background
\refs{\ts,\tfol}. Like the flat ten-dimensional GS action, the 
$AdS_5\times S^5$ action has four-dimensional and six-dimensional
classical
analogs -- the $AdS_2\times S^2$ and the $AdS_3\times S^3$ actions
respectively \refs{\tftwo,\tfthree}.  Unfortunately, it is not known
how to
quantize any of these GS actions.

Another conventional approach to constructing superstring actions uses
the
RNS formalism where quantization is straightforward since the action is
free in a flat background. The RNS formalism was successful for
quantizing
the superstring in an $AdS_3\times S^3$ background with NS/NS flux
\ref\kut{ A. Giveon, D. Kutasov, and N. Seiberg, {\it Comments On
String
Theory On ${\rm AdS}_3$}, hep-th/9806194.}.  However, the RNS approach
has
not been used to describe the superstring in RR backgrounds because of
the
complicated nature of the RR vertex operator.

Over the last five years, an alternative approach to constructing
superstring actions has been developed which combines the advantages of
the
GS and RNS approaches
\ref\mehybrid{N. Berkovits, {\it The Ten-Dimensional
Green-Schwarz Superstring is a Twisted Neveu-Schwarz-Ramond String },
Nucl. Phys. B420 (1994) 332, hep-th/9308129\semi N. Berkovits, {\it A
New
Description Of The Superstring}, Jorge Swieca Summer School 1995, p.
490,
hep-th/9604123.}.
Like the GS approach, this hybrid approach uses spacetime spinor
variables
as fundamental fields, allowing simple vertex operators for RR fields.
And
like the RNS approach, it reduces to a free action for a flat
background,
so quantization is straightforward. The worldsheet variables of the
hybrid
formalism are related to the RNS worldsheet variables by a field
redefinition, and the action contains critical $N=2$ worldsheet
superconformal invariance which replaces the $\kappa$-symmetry of the
GS
action. This $N=2$ worldsheet superconformal invariance is related to a
twisted BRST invariance of the RNS formalism and is crucial for
removing
unphysical states.

\nref\ufive{ N. Berkovits, {\it Quantization
of the Superstring with Manifest U(5) Super-Poincar\'e Invariance},
hep-th/9902099.}
\nref\topo{ N. Berkovits and C. Vafa, {\it $N=4$
Topological Strings}, Nucl. Phys. B433 (1995) 123, hep-th/9407190.}
\nref\mefour{N. Berkovits,  {\it Covariant Quantization of
the Green-Schwarz Superstring In A Calabi-Yau Background},
hep-th/9404162,
Nucl. Phys. {\bf B431} (1994) 258.}
\nref\siegel{N. Berkovits and W. Siegel,
{\it Superspace Effective Actions for 4D Compactifications of Heterotic and
Type II Superstrings}, hep-th/9510106, Nucl. Phys.  B462 (1996) 213.}

The only disadvantage of this hybrid approach is that ten-dimensional
Lorentz invariance cannot be kept manifest.  The maximum amount of
invariance which can be kept manifest (after Wick-rotating) is a U(5)
subgroup of the Lorentz group \ufive.  However, depending on the
desired
background, there are other ways of breaking the manifest Lorentz
invariance which are more convenient.  For example, one choice is to
break
the SO(9,1) Lorentz invariance down to $SO(5,1) \times U(2)$ \topo.
This
choice is convenient for describing compactifications of the
superstring to
six dimensions and was used succesfully in \bvw\ (see also \bzv) for
quantizing the superstring in an $AdS_3\times S^3$ background with RR
flux.
The worldsheet description of strings propagating on $AdS_3 \times S^3
$ is
given by a certain modification of a sigma model on $PSU(1,1|2)$.
Another
choice is to break the SO(9,1) Lorentz-invariance down to $SO(3,1)
\times
U(3)$ \refs{\mefour, \siegel}.  This choice is convenient for
describing
compactifications of the superstring to four dimensions and will be
used in
this paper for quantizing the superstring in an $AdS_2\times S^2$
background with RR flux.  The sigma model presented here is based on
the
coset supermanifold $PSU(1,1|2)/U(1) \times U(1)$.

\bigskip

As was shown in \bvw\ and \bzv, the $PSU(1,1|2)$ sigma model is
conformal
invariant because of the Ricci flatness of the group supermanifold.  To
get
an $AdS_2\times S^2$ model, one has to form a coset space by dividing
$PSU(1,1|2)$ by a $U(1) \times U(1)$ subgroup.  Unfortunately, the
resulting coset space turns out not to be Ricci flat. Indeed, on
general
grounds, only division by symmetric subgroups\foot{$H$ is a symmetric
subgroup of $G$ if it is the invariant locus of a ${\bf Z}_2$
automorphism
of $G$.}  preserves Ricci flatness.  Nevertheless, the $U(1)\times
U(1)$
subgroup is quite special -- it is the invariant locus of a $\Z$
automorphism of $PSU(1,1|2)$. It is also a symmetric subgroup of the
bosonic part of $PSU(1,1|2)$.  Thanks to this $\Z$ action, one can add
a WZ
term that can be used to restore one-loop conformal invariance.  This
WZ
term is $d$-exact and the corresponding interaction can be written in
terms
of manifestly $PSU(1,1|2)$ left invariant currents.  This mechanism for
constructing a conformal field theory works for any coset space $G/H$,
provided that $G$ is Ricci flat and $H$ is the invariant locus of a
$\Z$
automorphism of $G$.  For example, superstring theory on $AdS_3 \times
S^3$
can be obtained from a conformal field theory based on the $PSU(1,1|2)
\times PSU(2|2)/SU(2) \times SU(2)$ coset\foot {The details of this
construction and its relation to \bvw\ will be discussed in
\ref\me {N. Berkovits, {\it Quantization of the Type II Superstring
in a Curved Six-Dimensional Background}, hep-th/9908041.}.}.
Our construction also works for the $PSU(2,2|4)/SO(4,1) \times SO(5)$
coset
manifold and leads to a conformal field theory that could be the
starting
point for quantizing superstring theory on $AdS_5\times S^5$.

\bigskip

The $AdS_2\times S^2$ background appears as the near horizon limit of a
four-dimensional extremal black hole. To realize this background in
type
IIB string theory one has to consider compactification on a Calabi-Yau
(CY)
manifold $X$ and wrap an appropriate number of 3-branes over 3-cycles
of
$X$.  The four-dimensional metric would be that of an extremal black
hole.
The ten dimensional geometry is not a direct product and the complex
moduli
of the CY vary as a function of the radial coordinate 
of the black
hole.  In the near horizon limit the CY moduli are fixed by the
attractor
equation
\ref\attr{S. Ferrara, R. Kallosh and A. Strominger,
{\it N=2 Extremal Black Holes}, hep-th/9508072, 
Phys. Rev. {\bf D52} (1995) 5412.} 
and $X \to X_{attr}$. 
At the attractor point the periods $(p^I, q_I)$ of the CY are
proportional
to D-brane charges
\eqn\attr{{\rm Im}(Cp^I)= Q^I~,~~{\rm Im}(C q_I)={\tilde Q}_I~,}
where $Q^I$ and ${\tilde Q}_I$ are electric and magnetic charges and $C$ is
a complex constant.  The CY $X_{attr}$ is uniquely fixed by the D-brane
charges \attr, up to the complex constant $C$
\foot{This complex constant will be related in section (2.3) 
to the vacuum value of the supergravity vector compensator.}.
The radius of $AdS_2 \times S^2$ is given by
\eqn\rad{R={1 \over 4} \sqrt{Q^I Q^J N_{IJ}} }
where $N_{IJ}$ is the vector superpotential
\ref\astr{A. Strominger, {\it Macroscopic Entropy of $N=2$ Extremal
Black Holes}, Phys. Lett {\bf B 383} (1996) 39}.
The conformal theory describing the strings in the near horizon
geometry
($AdS_2 \times S^2$ background) factorizes into the product
\eqn\dirprod{ \Big( AdS_2 \times S^2 \Big) \times X_{attr} \times
H_{\rho}~,}
where the first factor is the conformal field theory of $AdS_2 \times
S^2$
constructed in this paper, the second factor is a sigma model on the
Calabi-Yau $X_{attr}$ (which can in principle be replaced by any
internal
$N=2$ $c=9$ superconformal theory), and the last factor is the
conformal
theory of a free chiral boson $\rho$.  The product \dirprod\ is not
exactly
a direct product since changing the D-brane charges adjusts the CY
moduli
fixed by \attr\ and at the same time changes the radius of $AdS_2
\times
S^2$.  This adjustment is ensured by the presence of manifest four
dimensional $N=2$ supersymmetry in our model.  The D-brane charges are
quantized, but this is a non-perturbative effect that cannot be seen in
the
perturbative worldsheet theory.

\bigskip

The plan of this paper is as follows: In section 2, we shall review the
four-dimensional version of the hybrid action in a flat \mefour\ and
curved
\siegel\ background, and then discuss this action in an $AdS_2\times
S^2$
background with RR flux. In sections 3-4, we shall show that this
hybrid
action is equivalent to a sigma model action for the coset
supermanifold
$PSU(1,1|2)/U(1)\times U(1)$ including a Wess-Zumino term.  This action
is
similar to the GS action considered by Zhou \tftwo (which was based on
the
$AdS_5\times S^5$ action of Metsaev and Tseytlin \ts), but has a
crucial
difference -- it includes a metric for the fermionic currents.  The
$\kappa$-symmetry is explicitly broken by the kinetic term for fermions
and
is replaced by $N=2$ worldsheet superconformal invariance.  At the end
of
section 4, we compute the one-loop beta functions and show that they
vanish for a certain coefficient in front of the WZ term.  We
demonstrate
that the sigma model on the coset space $G/H$ with the WZ term is
one-loop
conformal invariant provided that the group $G$ is Ricci flat and $H$
is
the fixed locus of a $\Z$ automorphism of $G$.

In section 5, we present a geometrical proof of one-loop conformal
invariance based on target-space considerations.  The target space of
our
sigma model is the coset supermanifold $G/H$ with a $G$-invariant
background metric $g_{AB}$ and an invariant antisymmetric tensor field
strenth $H_{ABC}$.  The choice of WZ term (i.e. antisymmetric field
strength $H$) is a subtle modification of the usual one which is
possible
because of $\Z$ symmetry.  After computing the curvature of super coset
spaces, we show that the non-vanishing Ricci curvature of the coset
space
is cancelled by the $H_{ABC}$ stress tensor in the appropriate
Einstein's
equations. Moreover, $H_{ABC}$ satisfies its own field equation.  In
addition, we confirm that the dilaton expectation value is constant to
two
loops because the coset supermanifolds in question have vanishing {\it
scalar} curvature.

Finally, in section 6 we speculate on the possible relation between
conformal field theories on the coset spaces $PSU(1,1|2)\times
PSU(2|2)/SU(2)\times SU(2)$ and $PSU(2,2|4)/SO(4,1)\times SO(5)$ and
quantization of the superstring in $AdS_3\times S^3$ and $AdS_5\times
S^5$
backgrounds.

\newsec{Hybrid Superstring in Four Dimensions}

We will start by describing the hybrid action based on the
$SO(3,1)\times
U(3)$ splitting of the Lorentz group \refs{\mefour,\siegel}.  This
action
is very similar to the four-dimensional version of the GS action, but
includes some crucial additional terms.  As will be reviewed below, the
Type II four-dimensional GS action contains four $\kappa$-symmetries
which
are replaced by $N=(2,2)$ superconformal invariance in the hybrid
action.
There are three main differences, however, between the four-dimensional
GS
action and the action based on the hybrid approach.  Firstly, the
action in
the hybrid approach is a free action in a flat background, so
quantization
is straightforward. Secondly, it includes compactification fields which
cancel the conformal anomaly. And thirdly, the action contains a term
coupling the dilaton zero mode with the worldsheet curvature, so
scattering
amplitudes have the expected dependence on the coupling constant.

\subsec{Hybrid approach in a flat four-dimensional background }

Before discussing the $AdS_2\times S^2 $ background, it will be useful
to
review the action in the hybrid approach for a flat four-dimensional
background. In the $SO(3,1)\times U(3)$ version there are ten bosonic
spacetime variables which split into $X^m$ for $m=0$ to 3, $Y^j$ and
$\bar
Y_j$ for $j=1$ to 3.  The $Y^j$ and $\bar Y_j$ variables describe the
compactification manifold while the $X^m$ variables describe the
four-dimensional spacetime.  As in the RNS approach, the hybrid model
has
left-moving fermions $(\psi^j,\bar \psi_j)$ and
right-moving fermions $(\widehat\psi^j,\widehat{\bar \psi}_j)$
associated
with  $Y^j$ and $\bar Y_j$.  
One also has sixteen fermionic worldsheet variables
transforming as four-dimensional spinors which split into $(\theta^\a,
p_\a)$, $(\bar\theta^\ad, \bar p_\ad)$ $({\widehat\theta}^\a, \widehat
p_\a)$, $({\widehat{\bar\theta}}{}^{\ad}, {\widehat{\bar p}}_{\ad})$
for
$\a, \ad=(1,2)~$\foot{The fermionic 
directions are labeled by $[\a, \ad,\ah,\adh]$ 
but in order to simplify the notation (in those cases where there is no
confusion), we will put a $hat$ only on top of the fields $\theta, \bar
\theta, p, \bar p,...~$.}.
The $\theta$'s and $\widehat\theta$'s correspond to the left-moving and
right-moving fermionic variables of $N=2$, $D=4$ superspace and the
$p\,$'s
and $\widehat p\,$'s are their conjugate momenta.  Finally, one has a
chiral and anti-chiral boson which will be called $\rho$ and
$\widehat\rho$.  One can show that the fermions and chiral boson of the
hybrid formalism are related by a field-redefinition to the ten
$\psi$'s,
two bosonic ghosts and two fermionic ghosts of the RNS formalism.  The
six
RNS $\psi$'s from the compactification directions are not quite the
same as
the six $\psi$'s in the hybrid formalism, but are related by a factor
involving the $\beta,\gamma$ ghosts.

In a flat or toroidal background, the worldsheet action for these
fields in
superconformal gauge is
\eqn\aab{\eqalign{
S &={1\over{\a '}}
\int dz d{\bar z}\,\, \Bigl[\,\half\dzp X^m \dzm X_m + p_\a \dzp\t^\a +
\bar p_\ad \dzp\tb^\ad  \cr
& \qquad\qquad\qquad +\widehat p_\a \dzm \th ^\a +
\widehat{\bar p}_\ad
\dzm\tbh{}^\ad
+ \dzp Y^j \dzm \bar Y_j + \psi^j \dzp \bar\psi_j
+ \widehat\psi^j \dzp \widehat{\bar\psi}_j \Bigr]~,}
}
where we have not tried to write the action for the chiral and
anti-chiral
boson $\rho$ and $\hat\rho$. The action of \aab\ is quadratic so all
fields
are free and it is completely straightforward to compute their OPE's.
We
only consider the Type II superstring in this paper, so all worldsheet
fields satisfy periodic boundary conditions.

The above action is manifestly conformal invariant, and it also
contains a
non-manifest $N=(2,2)$ superconformal invariance.  The left-moving
$N=2$
generators are given by
\eqn\aac{\eqalign{
T &=\half\dzm X^m  \dzm X_m +
p_\a\dzm \t^\a + \bar p_\ad \dzm\tb^\ad +{1\over 2}
\dzm\rho\dzm\rho  \cr
&\quad +\dzm Y^j\dzm \bar Y_j +\half(\psi^j\dzm\bar \psi_j
+\bar\psi_j\dzm \psi^j)~, \cr
G &=e^{i\rho} (d\,)^2  +\psi^j\dzm \bar Y_j\,, \qquad
\bar G  =e^{-i\rho} (\bar d\,)^2 ~+\bar \psi_j \dzm Y^j,\cr
J &=i \dzm\rho~+\psi^j\bar\psi_j~,}
}
where
\eqn\aad{\eqalign{
d_\a=p_\a+i\s^m_{\a\ad}\tba\dzm X_m -\half(\tb)^2\dzm\t_\a
+{1\over 4}\t_\a \dzm (\tb)^2, \cr
 \bar d_\ad=\bar p_\ad
+i\s^m_{\a\ad}\ta\dzm X_m -\half(\t)^2\dzm\tb_\ad
+{1\over 4}\tb_\ad \dzm (\t)^2~,}
}
and $(d)^2$ means $\half\epsilon^{\a\b} d_\a d_\b$. The right-moving
$N=2$
generators are obtained from \aac\ by replacing $\dzm$ with $\dzp$ and
placing hats on the worldsheet variables. Note that $(d_\a,\bar d_\ad,
\widehat d_\a, \widehat{\bar d}_\ad)$ anti-commute
with the spacetime-supersymmetry generators
\eqn\aai{\eqalign{
q_\a =\int dz \,Q_\a~ ~~{\rm where}~~~Q_\a=p_\a -i\s^m_{\a\ad}\tba\dzm
X_m-
        {1\over 4}(\tb)^2\dzm\t_\a~,\cr
\bar q_\ad=\int dz \, \bar Q_{\ad}~~~{\rm where}~~~
 \bar Q_{\ad}= \bar p_\ad
        -i\s^m_{\a\ad}\ta\dzm X_m-{1\over 4}(\t)^2\dzm\tb_\ad~.}
}
To obtain ${\widehat q}_\a$ and ${\widehat {\bar q}}_\ad$, one just
substitutes $\t \leftrightarrow \th$ and replaces $\partial$ with $\bar
\partial$.

The $N=2$ generators of \aac\ split into two pieces, one piece
depending on
the `six-dimensional' variables $(Y^j,\psi^j)$, and the other piece
depending on the remaining `four-dimensional' variables
$(X^m,\theta^\a,
\bar\theta^{\ad}, p_\a,\bar p_{\ad},\rho)$. One can check that the
six-dimensional contribution forms an $N=2$ $c=9$ algebra while the
four-dimensional contribution forms an $N=2$ $c=-3$ algebra, summing to
a
critical $N=2$ $c=6$ algebra.  The above action is easily generalized
to
the case when the six-dimensional compactification manifold is
described by
an $N=2$ $c=9$ superconformal field theory.  In this case, one simply
replaces the flat six-dimensional contribution to the action and $N=2$
generators by their non-flat six-dimensional counterpart.

The action of \aab\ can be written in manifestly
spacetime-supersymmetric
notation using the supersymmetric combinations
\eqn\pidef{\Pi^m_j=\partial _j X^m +i\s^m_{\a\ad}
(\tb^\ad \partial _j \t^\a +\t^\a \partial _j \tb^\ad+
\tbh{}^\ad \partial _j \th{}^\a +\th{}^\a \partial _j \tbh{}^\ad).}
In terms of $\Pi^m _j$, the action \aab\ can be written as a sum of the
action describing an $N=2$ $c=9$ superconformal theory $S_C$ and a
four-dimensional contribution
\eqn\susyflat{\eqalign{  
S &= S_C + {1\over{\a '}}
\int dz d \bar z  \,\,\Bigl( \,\half\Pi^m _z \,\Pi_{\bar zm} +
d_\a \Pi^{\a }_{\bar z} +
\bar d_\ad \Pi^{\ad }_{\bar z} + {\widehat d}_\a {\widehat \Pi}_z ^{\a}
+
\widehat{\bar d}_\ad {\widehat \Pi}_z ^{\ad}  \cr
&\qquad\qquad\qquad + \epsilon^{jk}\,\bigl[\, \Pi^m_j
i\s_m^{\a\ad}(\t_\a\p_k\tb_\ad
+\tb_\ad\p_k\t_\a
-\th_\a\p_k\tbh_\ad
-\tbh_\ad \p_k\th_\a)  \cr
&\qquad\qquad\qquad +(\t_\a\p_j\tb_\ad
+\tb_\ad\p_j\t_\a)
(\th{}^\a \p_k\tbh{}^\ad
+\tbh{}^\ad \p_k\th^\a)\bigr]  \Bigr) }
}
where $j,k=z, \bar z$ and $d_\a$ differs from the definition of \aad\
by
terms which vanish on-shell.  Note that the last two lines of
\susyflat\
represent the standard Wess-Zumino term of the four-dimensional Type II
Green-Schwarz action in conformal gauge.

\subsec{Type II action in a general curved background}

Using the action of \aab\ in a flat background and the massless vertex
operators described in \siegel, it is easy to guess the following
action in
a curved background:
\eqn\alb{
\eqalign{
S &=  S_C + {1\over{\a '}}
\int d \bar z  dz \,\,\Bigl( \,\half\Pi^c _{\bar z}\,\, \Pi_{zc} + 
B_{AB}\, \Pi^A _z\, \Pi^B _{\bar z} \cr
&\qquad\qquad  +d_\a \Pi^{\a }_{\bar z} +
\bar d_\ad \Pi^{\ad }_{\bar z} + 
\widehat d_\a {\widehat \Pi}_z ^{\a} +
\widehat{\bar d}_\ad ~{\widehat \Pi}_z ^{\ad}  + \cr
&\qquad\qquad +d_\a P^{\a\bh} ~\widehat d_\b +
\bar d_\ad \bar P^{\ad\bdh} ~ \widehat{\bar d}_\bd +
d_\a Q^{\a\bdh} ~\widehat{\bar d}_\bd +
\bar d_\ad\bar Q^{\ad\bh}~ \widehat d_\b \Bigr)  ~,}
}
where $\Pi^A _j$ is written in terms of the supervierbein $E_M {}^A$ as
$\Pi^A_j=E_M{}^A \partial _j Z^M$ with $Z^M=(X^m,~
\t^\mu,~\tb^{\dot\mu},
~\th ~^\mh, ~\tbh ~^\mdh)$, and the index $A$ takes the
tangent-superspace
values [$c,\alpha,\ad,\ah,\adh$].  Once again, to simplify notation we
denote $\t^{\ah}$ as ${\widehat \t}^{\a}$ and $\t ^{\adh}$ as
${\widehat
\t}^{\ad}$.  In other words, we put the $hat$ over the symbol instead
of
the index.  We will also use this rule for all other worldsheet fields.
But we will use the standard notations [$c,\alpha,\ad,\ah,\adh$] for
the
background fields such as $Q^{\a \bh}$.
 
Spinor indices [$\a$,$\ad$] and [$\ah$,$\adh$] correspond to the left
and
right-moving degrees of freedom and $c$ is a vector index.  The lowest
component of $E_m^c$ is the vierbein and the lowest components of
$E_m^\alpha$ and $E_m^{\ah}$ are the gravitini. The antisymmetric field
$B_{MN}$ is the two-form superfield where the lowest component of
$B_{mn}$
is the NS/NS two-form.  The superfields $P^{\a\bh}$ and $P^{\ad\bdh}$
contain the self-dual and anti-self-dual parts of the RR graviphoton
field-strength as their lowest components.  Similarly, the superfields
$Q^{\a\bdh}$ and $\bar Q^{\ad\bh}$ contain the derivatives of the two
RR
scalars as their lowest components.  The last term in \alb, $S_C$, is
the
action that describes the compactification manifold.

In a flat background, one can check that  
the action \alb\ reduces to \susyflat.
Furthermore, linear perturbations around the flat background
reproduce the massless vertex operators of \siegel \foot{As explained 
in reference \siegel, 
the complete Type II action also contains a `Fradkin-Tseytlin' term
which
couples the spacetime dilaton to the $N=(2,2)$ worldsheet
supercurvature.
But since the dilaton is constant in the $AdS_2\times S^2$ background,
this
term will only give the contribution $\phi\int R + a\int F$ where $\int
R$
is the worldsheet Euler number, $\int F$ is the worldsheet U(1)
instanton
number (coming from the U(1) gauge field of the N=2 worldsheet), and
$e^{-\phi-i a}$ is the background value of the supergravity vector
compensator $Z^0$ (which is related by supergravity equations of motion
to
the dilaton and NS/NS axion background values) \siegel.}.

Note that $d$ and $\widehat d$ are fundamental fields in the action of
\alb, and in a flat background, they can be expressed in terms of the
free
fields $p$ and $\bar p$ using the definitions \aad.  If $d$ and
$\widehat
d$ are set to zero and the compactification fields are ignored, the
action
of \alb\ is precisely the four-dimensional version of the covariant GS
action in a curved background.  When the background satisfies certain
torsion constraints, the covariant GS action has $\kappa$-symmetries
which
allow half of the $\theta$'s to be gauge-fixed. However, it has only
been
possible to quantize this action in light-cone gauge, which is not
accessible for arbitrary backgrounds \ref\deboer{J. de Boer 
and K. Skenderis, {\it Covariant Computation
of the Low Energy Effective Action of the Heterotic Superstring},
Nucl. Phys. B481 (1996) 129.}.

By including the $d$ and $\widehat d$ fields, as well as the chiral
boson
$\rho$ and the compactification fields, it is possible to covariantly
quantize the action \alb\ in a manner analogous to the
normal-coordinate
expansion used in bosonic string or RNS superstring sigma models.  In a
curved background, one still has an $N=2$ superconformal invariance at
the
quantum level if the background fields are on-shell.  The $N=2$
left-moving
generators are given by
\eqn\aaf{\eqalign{
T &=  \Pi^c_z \Pi_{cz} + d_\a \Pi^\a_z +  {\bar d}_\ad 
\Pi^{\ad}_z
 +{1\over 2}
\dzm \rho \dzm \rho ~  +T_C~,\cr
G &=e^{i\rho} (d)^2 ~+G_C, \quad
\bar G=e^{-i\rho}  (\bar d)^2 ~+\bar G_C, \cr
J &=i \dzm\rho~+J_C~,}
}
where $[T_C,G_C,\bar G_C,J_C]$ are the $N=2$ $c=9$ generators
describing
the compactification manifold.  One can check that the classical
equations
of motion of \alb\ imply that $(d)^2$ and $(\bar d)^2$ are holomorphic,
so
the $N=2$ generators are holomorphic at least at the classical level
\refs{\siegel,\deboer}. Holomorphicity at the quantum level is expected
to
imply that the background superfields satisfy their low-energy
equations of
motion. This has been explicitly checked at one-loop for the heterotic
superstring in a curved background \deboer, but has not yet been
checked
for the Type II superstring in a curved background. So the action of
\alb\
is still a conjecture at the quantum level but we will provide evidence
for
this conjecture by explicitly showing that the one-loop conformal
anomaly
vanishes when the background is chosen to describe $AdS_2\times S^2$.

\subsec{Type IIA action in  $AdS_2\times S^2$ background with RR flux}

To describe the $AdS_2\times S^2$ background with RR flux, one simply
needs
to substitute the values for the superfields into \alb.  In this
background, the field-strength of the RR graviphoton is equal to
$F_{01} =
F_{23} = N$ where the integer $N$ counts the number of branes.
But in the presence of a Type IIA (or Type IIB) Calabi-Yau
compactification
with $h$ Kahler moduli (or $h$ complex moduli), there is some ambiguity
which of the $h+1$ RR vector fields is called the graviphoton. There is
an
auxiliary two-form, $T_{\mu\nu}$, in the supergravity multiplet whose
on-shell self-dual part satisfies the equation of motion
\ref\dewit{B. deWit, P.G. Lauwers and A. Van Proeyen, {\it Lagrangians
of
N=2 Supergravity-Matter Systems}, Nucl. Phys. B255 (1985) 569.}
\eqn\tdef{
T_{\a \bh} ={ {{4 N_{PQ} Z^Q}}\over {N_{RS} Z^R Z^S}}
F_{\a \bh}^P\,\,,
}
where $P,Q,R,S=0,...h$, $Z^0= e^{-\phi-ia}$ is the supergravity vector
compensator, $Z^P/Z^0$ are the Kahler (or complex) moduli for
$P=1,...h$,
$N_{PQ}$ is the vector superpotential, and $F_{\a\bh}^P$ is the
self-dual
part of the $P^{th}$ vector field strength.  $T_{\a\bh}$ is the lowest
component of the superfield $P_{\a\bh}$ and $ F_{\a\bh} = Z^0
T_{\a\bh}$
gives the linear combination of RR field-strengths which is turned on.
So
in the $AdS_2\times S^2$ background, $P^{\alpha\beta}$ and $\bar
P^{\ad\bd}$ take the values
\eqn\value{\eqalign{
P^{\a\bh} = (Z^0)^{-1} F^{\a\bh}
= (Z_0)^{-1} N(\sigma^{01})^{\a \bh} = (Z_0)^{-1} N\delta^{\a \bh}~,
\cr
\bar P^{\ad\bdh} =
(\bar Z^0)^{-1} F^{\ad \bdh} =
(\bar Z^0)^{-1} N(\sigma^{01})^{\ad \bdh} =
(\bar Z^0)^{-1} N\delta^{\ad \bdh}~.}
}
The dependence of $P^{\a\bh}$ on the string coupling constant
$g = e^\phi$ can be understood by recalling that the $e^{-2\phi}$ term
is
absent in the $FF$ kinetic term for the RR field.  So $P^{\a\bh}$ is
related to the RR field strength by a factor of $g$. The dependence of
$P^{\a \bh}~$ on the phase factor $e^{ia}$ from $(Z_0)^{-1}$ comes from
its
non-zero $R$-charge where $e^{ia}$ is related to the phase of the
constant
$C$ in \attr.  In the rest of this paper, we shall assume that $a=0$
although all our formulas can be generalized to non-zero $a$ by simply
rotating all fields by a phase factor proportional to their $R$-charge.

Note that the $h$ linear combinations of field-strengths which are {\it
not} turned on are given by
\eqn\notto{
F_{\a \bh}^P - {1\over 4} Z^P T_{\a\bh}\,\,,
}
for $P=1,...h$.  This can be seen from $N=2$ $D=4$
spacetime-supersymmetry
since the Calabi-Yau gauginos transform into \notto\ under
supersymmetry
\astr.

Substituting \value\ into \alb, we obtain
\eqn\amb{\eqalign{S &=  
{1\over{\a '}}
\int d \bar z  dz \Bigl(\,\half \Pi^c_{\bar z}\,\, \Pi_{zc} + 
B_{AB}\,\Pi^A _z \,\, \Pi^B _{\bar z}  \cr
&\qquad\qquad + d_\a \Pi^{\a }_{\bar z}  +
\bar d_\ad \Pi^{\ad }_{\bar z}
+
\widehat d_\a \,\widehat \Pi^{\a }_z +
\widehat{\bar d}_\ad \,\widehat \Pi^{\ad }_z \cr
&\qquad\qquad + 
Ng d_\a  \,\widehat d_\b \,\delta^{\a\b} +
Ng \bar d_\ad \widehat{\bar d}_\bd \,\delta^{\ad\bd}\Bigr) + S_C ~. }
}
In the above action, $\Pi^A$ and $B_{AB}$ are defined in precisely
the
same manner as in the covariant GS $AdS_2\times S^2$ action of Zhou
\tftwo,
which was based on the $AdS_5\times S^5$ action of Metsaev and Tseytlin
\ts.  However, the second and third lines of \amb\ are not present in
the
usual GS action and are crucial for quantization.  As discussed in
\tftwo\
and will be reviewed in the next section, the objects $\Pi^A$ can be
constructed out of currents of the supergroup $PSL(1,1|2)/U(1)\times
U(1)$.
In the $AdS_2\times S^2$ background, $B_{AB}$ can be written in the
following simple form:\foot{It appears to have never been mentioned in
previous papers \refs{\ts{--}\tfthree}\ that the WZ term for $AdS$
backgrounds is exact in the group-invariant sense, i.e. that the $B$
field
can be constructed out of the supersymmetric currents.  This simple
form
for $B_{AB}$ was originally found in 
\ref\green{M.B. Green, {\it Supertranslations, Superstrings and
Chern-Simons Forms}, Phys. Lett. 223B (1989) 157\semi W. Siegel, {\it
Randomizing the Superstring}, Phys. Rev. D50 (1994) 2799,
hep-th/9403144.}
which considered a version of the GS superstring with two types of
fermionic symmetries.}
\eqn\Bsimple{ 
B_{\alpha \bh} =B_{\bh\a}=  -{1\over {4Ng}}\d_{\a\bh},\quad
B_{\ad \bdh} =B_{\bdh \ad}= -{1\over {4Ng}}\d_{\ad\bdh}\,\,,
} 
with all other components of $B_{AB}$ vanishing.  A similarly simple
form
for $B_{AB}$ occurs in the $AdS_3\times S^3$ and $AdS_5\times S^5$
backgrounds. Equation \Bsimple\ is easy to prove since $H_{ABC} =
\nabla_{[A} B_{BC]} + T_{[AB}{}^D B_{C]D}$, so the only non-zero
components
of $H_{ABC}$ are
\eqn\Hsimple{\eqalign{
  H_{c \a\ad} &= T_{c \a}{}^{\bdh} B_{\ad \bdh} +
 T_{c \ad}{}^{\bh} B_{\a \bh}\,\, \cr 
&= Ng(\s_c)_{\a\gd}
\d^{\gd\bdh}(-{1\over{4Ng}}  \d_{\ad\bdh}) +
Ng (\s_c)_{\g\ad} \d^{\g\bh}
(-{1\over {4Ng}}\d_{\a\bh}) 
 =-\half(\s_c)_{\a\ad}\,\,, }
}
and $H_{c \ah\adh}=\half(\s_c)_{\ah\adh}$, 
which can be obtained from \Hsimple\ by replacing
the
fermionic unhatted indices by hatted and vice versa.  The precise
normalizations are fixed by the supergravity equations.  In \Hsimple,
we
have used the torsion constraints of \siegel\ to relate the torsion
components to the Ramond-Ramond
field-strengths by the equations
\eqn\tconst{\eqalign{
T_{c \a}{}^{\bdh} 
 = (\s_c)_{\a\gd}\,
\bar P^{\gd\bdh}, \quad
T_{c \ad}{}^{\bh} 
 =(\s_c)_{\g\ad}\,
P^{\g\bh}, \quad \cr
T_{c \ah}{}^{\bd} 
 = -(\s_c)_{\ah\gdh}\,
\bar P^{\bd\gdh}, \quad
 T_{c \adh}{}^{\b} 
 =-(\s_c)_{\gh\adh}\,
P^{\b\gh}
~~.}
} 

As in the hybrid action for $AdS_3\times S^3$ \bvw, it is convenient to
integrate out the $d$ and $\widehat d$ fields. This is possible because
of
the quadratic term $d\widehat d$ produced by the RR flux, which implies
that $d$ and $\widehat d$ are auxiliary fields.  Substituting the
equations
of motion for $d$ and $\widehat d$ into \amb\ and taking \Bsimple\ into
account, one obtains
\eqn\afx{\eqalign{S=S_C+
{1\over{ \a '}}
\int  dz d \bar z \, \Bigl[\, \half\Pi^c_{\bar z} \,\, \Pi_{zc} + 
&{3 \over {4Ng}}( \d_{\a \b} {\widehat\Pi}_z ^{\a} \Pi_{\bar z} ^{\b}+
\d_{\ad \bd} {\widehat\Pi}_z ^{\ad} \Pi_{\bar z} ^{\bd})+\cr
+&{1 \over {4Ng}}( \d_{\a \b} {\widehat\Pi}_{\bar z} ^{\a} \Pi_{z}
^{\b}+ 
\d_{\ad \bd} {\widehat\Pi}_{\bar z} ^{\ad} \Pi_{z} ^{\bd})\, \Bigr]~.}
}
The $Ng$ dependence can be simplified by rescaling $E^c_M\to 
(Ng)^{-1} E^c_M$, and $E^\a_M\to (Ng)^{-\half} E^\a_M$ to obtain\foot{This 
rescaling is chosen such that the torsion constraint 
for $T^c_{\a\ad}$ remains independent of $Ng$.}
\eqn\aff{\eqalign{S=S_C+
{1\over{ 2\a 'g^2 N^2}}
\int  dz d \bar z \, \Bigl[\,\Pi^c_{\bar z} \,\, \Pi_{zc} + 
&{3 \over {2}}( \d_{\a \b} {\widehat\Pi}_z ^{\a} \Pi_{\bar z} ^{\b}+ 
\d_{\ad \bd} {\widehat\Pi}_z ^{\ad} \Pi_{\bar z} ^{\bd})+\cr
+&{1 \over {2}}( \d_{\a \b} {\widehat\Pi}_{\bar z} ^{\a} \Pi_{z}
^{\b}+ 
\d_{\ad \bd} {\widehat\Pi}_{\bar z} ^{\ad} \Pi_{z} ^{\bd})\, \Bigr]~.}
}

The $N=2$ superconformal generators are the same as in \aaf\ with $d$
replaced by its equation of motion, and the
compactification-independent
part of the stress tensor will be shown in section 3 to be the
conformal
generator for a sigma model based on the supergroup
$PSU(1,1|2)/U(1)\times
U(1)$.  Note that the action and constraints are invariant under the
transformation which takes the bosonic currents $\Pi ^c \to - \Pi ^c$,
the fermionic currents $\Pi \to i \Pi $ and ${\widehat
\Pi}{} \to -i {\widehat \Pi}$, and which shifts $\rho \to\rho+\pi$ and
$\bar\rho \to\bar\rho+\pi$.  The presence of this $\Z$ symmetry will be
important later.

\subsec{Perturbative derivation of $AdS_2\times S^2$ action}

In the hybrid action for $AdS_3\times S^3$ of reference \bvw, the
action
was justified using a Ramond-Ramond vertex operator to perturb around a
flat background. This justification was important in the $AdS_3\times
S^3$
case since there did not exist a hybrid action for the general
six-dimensional Type II superstring background. Although the existence
of
\alb\ as a hybrid action for the general four-dimensional background
\siegel\ makes such a justification less necessary in the $AdS_2\times
S^2$
action, we shall show in this section that such a perturbation is
possible
and leads to the same result of \afx.

The vertex operator in a flat background for a constant graviphoton
field-strength $F^{01}=F^{23}=N$ is given by
\eqn\vergrav{
V=Ng\int dz d \bar z
\,(\d_{\a\b} Q^\a \widehat Q^\b +\d_{\ad\bd}\bar Q^{\ad} 
\widehat {\bar Q}{}^{\bd})\,\,,
}
where the $g$ dependence in $V$ comes for the same reason as in \value\
and
$Q$ is given in terms of $p$'s and $\theta$'s by \aai.  It is clear
that
$V$ is a physical operator since the supersymmetry currents $Q$ commute
with the $N=2$ constraints of \aac.
Adding $V$ to the flat action of \aab\ produces the action
\eqn\flatads{\eqalign{ 
S & = { 1\over{\a '}}
\int dz d \bar z  \,\,\Bigl( \,\half\dzp X^m \dzm X_m + p_\a \dzp\t^\a
+
\bar p_\ad \dzp\tb^\ad   \cr
&\qquad \qquad + \widehat p_\a \dzm \th^\a +
\widehat{\bar p}_\ad
\dzm \tbh{}^\ad  +
Ng(\d_{\a\b} Q^\a \widehat Q^\b +\d_{\ad\bd}\bar Q^{\ad} 
\widehat{\bar Q}{}^{\bd})\Bigr) + S_C~.}
}
Now one can integrate the $Q$'s out, and keeping terms only up to the
cubic
order in fields, one can just replace $Q=(Ng)^{-1} \partial \th$ and
$\widehat Q=-(Ng)^{-1} \bar \partial \t$.  Finally, integrating by
parts
and using the fact that terms proportional to $\dzm \dzp X^m$ or $\dzm
\dzp
\t$ can be removed by redefining $X^m$ or $\t$, one can write the
action as
\eqn\finalflatads{\eqalign{ 
S &=  {1\over{\a '}}
\int dz d \bar z \,\,\Bigl( \,\half\dzp X^m \dzm X_m + 
(Ng)^{-1} (\d_{\a\b} \dzm \th^\a \dzp \t^\b + 
          \d_{\ad\bd} \dzm \tbh {}^{\ad} \dzp \tb^{\bd})  \cr
&\qquad +  i\s^m_{\a\ad}
\,(\tb^{\ad} \p_j X_m \p_k \t^\a +
\t^{\a} \p_j X_m \p_k \tb^\ad -
\tbh {}^{\ad} \p_j X_m \p_k \th {}^\a -
\th {}^{\a} \p_j X_m \p_k \tbh {}^\ad
)\Bigr)  + S_C\,.} }
The action \finalflatads\ supplemented by the total derivative term
\eqn\extr{
\Delta L=- \epsilon^{jk} {1 \over {2Ng\a'}}\, \Bigl(  
\d_{\a\b} \p_j \th^\a \p_k \t^\b + 
\d_{\ad\bd} \p_j \tbh {}^\ad \p_k \tb^\bd
\Bigr)~, }
reproduces the action of \afx\ up to cubic order in the fields.

\newsec{$AdS_2 \times S^2$ from the $PSU(1,1|2)/U(1)\times U(1)$ coset}

{}From the work of \bvw\ and \bzv, it is natural to expect that 
quantization of strings in the $AdS_2 \times S^2$ background requires a
sigma model based on a quotient supermanifold $PSU(1,1|2)/U(1) \times
U(1)$.  In this section, we discuss this coset supermanifold and its
higher
dimensional analogs. We also introduce a $\Z$ symmetry that will play
an
important role in our considerations.  Finally, we construct a sigma
model
action as a gauged principal chiral field. The model constructed in
this
section is not conformal and the necessary modification (a WZ term)
will be
discussed in the next section.

As we will confirm in this and in the next sections, the action \aff\
indeed combines the sigma model and the WZ terms of the
$PSU(1,1|2)/U(1)
\times U(1)$ coset.  To make the identification, one notices that the
objects $[\Pi^{c}, \Pi^{\a}, \Pi^{\ad}, \widehat \Pi ^{\a},\widehat \Pi
^{\ad}]$ generate global $PSU(1,1|2)$ rotations and can be identified
with
the sigma model currents.

\subsec{Coset spaces and $\Z$ symmetry}

Let us start this section with a discussion of the signature of the
coset
space. There are two closely related cosets, $PSU(2|2)/U(1) \times
U(1)$
and $PSU(1,1|2)/U(1) \times U(1)$, that differ by the choice of real
structure on the group. The first coset is based on the $SU(2)\times
SU(2)
\subset PSU(2|2)$ and leads to the $S^2 \times S^2$ geometry with the
signature being $(2,2)$. The other coset leads to the $AdS_2 \times
S^2$
geometry with the signature $(1,3)$.  Clearly, the physics of these two
backgrounds is very different, but they share many common algebraic
properties.

The super Lie algebras $psu(2|2),~psu(1,1|2)$ are the algebras of $4\times
4$ matrices with bosonic diagonal blocks and fermionic off-diagonal blocks
\eqn\gr{ M=\pmatrix{ A & X \cr  Y & B} ~~~{\rm where}~~tr A = tr B =
0~.}
The bracket is defined to be the commutator projected on the
doubly-traceless subspace. The supertrace is defined as $Str(M)=TrA-TrB$
and the (super) antihermiticity condition for $psu(2|2)$ is
\ref\kac{V. Kac, {\it A Sketch of Lie Superalgebra Theory}, 
Comm. Math. Phys. 53 (1977) 31.}
\eqn\her{ M^\dagger \equiv \pmatrix{ A^\dagger  & -iY^\dagger \cr  
-iX^\dagger &
B^\dagger} = -M \quad \to\quad
A=-A^{\dagger}~,~~B=-B^{\dagger}~,~~X=iY^{\dagger}.}
In other words, the Grassmann even matrices $A,B$ are traceless
antihermitian, and the Grassmann odd $X,Y$ matrices are related to each
other.\foot{Antihermitian (super)matrices form a superalgebra because
$(MN)^\dagger = N^\dagger M^\dagger$. On block matrices dagger means
transposition composed with ordinary complex conjugation (
$\overline{(\epsilon_1\epsilon_2)} = \bar\epsilon_1
\bar\epsilon_2$). } For the case of $psu(1,1|2)$, the
anti-hermiticity condition is
\eqn\heroth{
M^\dagger \equiv \pmatrix{ \sigma_3A^\dagger \sigma_3 & -i\sigma_3
Y^\dagger \cr   -iX^\dagger \sigma_3 &
B^\dagger} = -M \,\, \to\,\, A=-\sigma_3 A^{\dagger} \sigma_3 ^{-1}
~,~~B=-B^{\dagger}~,~~X=i \sigma_3 Y^{\dagger}.}

The bosonic geometry of the $PSU(1,1|2)$ is $AdS_3 \times S^3$, so to
construct a string theory on $AdS_2 \times S^2$, we have to quotient the
group by (the right action of) $U(1) \times U(1)$.  The first $U(1)$ is
embedded into $SU(1,1)$ to produce $AdS_2$, while the second is embedded
into $SU(2)$ to give rise to $S^2$.  Since $PSU(1,1|2)$ has already the
right number of fermions (8 real ones), we divide only by a bosonic
subgroup.  For a reason that will become clear later, we will think of this
subgroup as an invariant locus of a $\Z$ automorphism.  Both for
$psu(1,1|2)$ and $psu(2|2)$ the $\Z$ automorphism is generated by
conjugation $M\to \Omega(M) \equiv {\Omega}^{-1} M \Omega$ with the matrix
\eqn\con{ \Omega= \pmatrix{ \sigma_3 & 0 \cr
  0 & i \sigma_3 }~.}  
This conjugation respects the anti-hermiticity conditions given
above and manifestly gives an algebra automorphism. In addition, the
invariant subalgebra $\Omega(M) = M$ is the desired bosonic $u(1)\oplus
u(1)$ algebra. Finally, $\Omega^4 (M) = M$.

Since this $\Z$ automorphism will play a key role in our construction, it
is of interest to show that it is also present for other $AdS_d \times S^d$
spaces of relevance.  For example, $AdS_5 \times S^5$ appears as the
bosonic part of the super-quotient $PSU(2,2|4)/SO(4,1) \times SO(5)$.  In
general, the Lie superalgebra $psu(n,n|2n)$ has a $\Z$ automorphism whose
invariant locus is $usp(n,n) \times usp(2n)$
\ref\serganova{V. Serganova, private communication.}. 
The anti-hermiticity condition for $psu(n,n|2n)$ is as follows
\eqn\genant{
M^\dagger \equiv \pmatrix{ \Sigma A^\dagger \Sigma & -i \Sigma
Y^\dagger \cr   -iX^\dagger \Sigma &
B^\dagger} = -M \,\, \to\,\, A=-\Sigma A^{\dagger} \Sigma
~,~~B=-B^{\dagger}~,~~X=i \Sigma Y^{\dagger}~,}
where $\Sigma=\sigma_3 \otimes I_n$ and $I_n$ is $(n \times n)$ identity
matrix.  Then the $\Z$ automorphism is generated by
\eqn\zact{ M = \pmatrix{ A & X \cr Y & B } \to \Omega(M) \equiv
\pmatrix{ J A^t J & -J Y^t J \cr J X^t J & J B^t J}\,,
\qquad {\rm where} \quad J=\pmatrix{ 0 & -I_{n} \cr I_{n} & 0 }~.}
While it is not expressed as conjugation, one can verify that it is a Lie
algebra automorphism compatible with the antihermiticity condition
\genant. Moreover, its fourth power is the identity and the $\Z$ invariant 
locus is precisely $usp(n,n) \times usp(2n)$.
When $n=2$,
$su(2,2) \sim so(4,2)$, $usp(2,2) \sim so(4,1)$, $usp(4) \sim so(5)$
(see
\ref\gil{R. Gilmore,  {\it Lie Groups, Lie Algebras and some of their
applications.},  John Wiley and Sons (1974)}) and we recover the
quotient that leads to $AdS_5 \times S^5$ geometry. 
Other group signatures can be discussed similarly\foot{Let us also 
mention that there is an 
alternative way to impose a hermiticity condition based on a second  
way to define complex
conjugation on Grassmann variables. This conjugation, denoted $\#$, 
satisfies \ref\ritt{V. Rittenberg and M. Scheunert, {\it Elementary
construction of graded Lie groups}, J. Math. Phys. 19 (1978)
709;\hfill\break
L. Frappat, P. Sorba, and A. Sciarrino, {\it Dictionary on Lie
Superalgebras},
hep-th/9607161.}:~$(c \epsilon)^\# = \bar c \, \epsilon^\#$,  
 $(\epsilon ^\# ) ^{\#} = (-)^{{\rm deg}(\epsilon)} \epsilon$, 
and 
$(\epsilon_1 \epsilon_2)^{\#} = \epsilon_1^\# \epsilon_2^\#$.
The alternative antihermiticity condition is 
\eqn\herr{ M^+ \equiv \pmatrix{ A^+  & Y^+ \cr  
-X^+ &
B^+} = -M \quad \to\quad
A=-A^+~,~~B=-B^+~,~~X=-Y^+\, }
where on block matrices ${}^+$ denotes transposition composed with
$\#$-conjugation. Note that $(MN)^+ = N^+ M^+$.  The above antihermiticity
condition is also compatible with the $\Z$ action \zact\ and restricts
$psl(2n|2n)$ (over the complex) to $psu(2n|2n)$ and the invariant locus to
$usp(2n) \times usp(2n)$ ($usp(4) \sim so(5)$ for $n=2$). }.

\medskip

The $\Z$ action can be used to decompose the Lie algebra ${\cal G}$ as
\eqn\dec{
{\cal G}=\H_0 \oplus \H_1 \oplus \H_2 \oplus \H_3~\,,}
where the subspace $\H_k$ is the eigenspace of the {\bf Z}$_4$ generator
$\Omega$ with eigenvalue $(i)^k$.  The subspaces $\H_1$ and $\H_3$ contain
all the fermionic generators of the algebra ${\cal G}$ and $\H_2$ contains
the bosonic generators of ${\cal G}$ that are not in the subalgebra $\H_0$.
The definition of {\bf Z}$_4$ implies that the hermitian conjugate of
$\H_1$ is $\H_3$ \foot{The eigenvectors in $\H_1$ and $\H_3$ thus give a
complex basis for the real Lie algebra ${\cal G}$.}.  Given that $\Z$ is an
automorphism of the Lie algebra, the decomposition $\oplus_k \H_k$
satisfies
\eqn\red{[\H_m, \H_n] \subset \H_{m+n}~ ({\rm mod~4})~.}
The bilinear form is also $\Z$ invariant and hence
\eqn\blue{
\langle \H_m, \H_n \rangle=0~~{\rm unless}~~~n+m=0~({\rm mod}~4)~.
}

To illustrate these ideas concretely, we present the basis for the
$psu(2|2)$
algebra explicitly and the
$\Z$ decomposition.  The Lie brackets of the algebra are
\eqn\commut{
\eqalign{  [K_{\mu\nu} , K_{\rho\sigma}] &= \delta_{\mu\rho}
K_{\nu\sigma}
-\delta_{\mu\sigma} K_{\nu \rho}  - \delta_{\nu \rho} K_{\mu \sigma}
+\delta_{\nu\sigma} K_{\mu \rho} ~,
\cr [K_{\mu\nu} , S_{\rho\alpha}] &=
\delta_{\mu\rho} S_{\nu\alpha}-\delta_{\nu\rho} S_{\mu\alpha} ~, \cr
\{ S_{\mu\alpha} , S_{\nu\beta}\} &= \half\,
\epsilon_{\mu\nu\rho\sigma}\,
\e_{\alpha\beta}\,K^{\rho\sigma}~, 
\cr}
}
where the indices $\mu, \nu, \rho, \sigma=(0,...3)$ and $\a, \b=(1,2)$.
The invariant bilinear form in the algebra is
\eqn\bilin{
\langle K_{\mu\nu}, K_{\rho\sigma} \rangle =
\epsilon_{\mu\nu\rho\sigma}\,, \qquad
\langle S_{\mu\alpha}\,, S_{\nu\beta} \rangle = \delta_{\mu\nu}
\epsilon_{\alpha\beta}.
}
Denoting $S_\mu \equiv S_{\mu1}\,, \widetilde S_\mu \equiv
S_{\mu2}$, one obtains for the $\Z$ invariant subspaces:
\eqn\spacesplit{
\eqalign{ \H_0  &= \{ K_{03},\, K_{12} \} \cr
\H_1  &= \{ S_\mu + i \widetilde S_\mu \bigl| ~\mu = 0, \cdots 3. \} \cr
\H_2  &= \{ K_{01},\, K_{02}, \,K_{13},\, K_{23} \} \cr
\H_3  &= \{ S_\mu  - i \widetilde S_\mu \bigl| ~\mu = 0, \cdots 3. \} \,. \cr }
}
The reader can explicitly check that this decomposition satisfies
\red\ and \blue.

\subsec{Sigma model action}

The easiest way to construct a sigma model action on the coset space is
by
gauging the subgroup $H$ whose Lie algebra is ${\cal H}_0$ (which will
sometimes be simply called $\cal H$).  Let $g(x) \in G$ describe the
map
from the worldsheet into the group $G$. The current $J=g^{-1}dg$ is
valued
in the Lie algebra ${\cal G}$.  Introducing the gauge field ${\cal A}$
taking values in $\H_0$, one can define a {\it gauged} action
\eqn\gauge{S[G,{\cal A}]={1 \over 4 \pi \lambda^2 } \int d^2 x ~
Str\big(J-{\cal A}  \big)^2 ~.}
It is convenient to decompose the current $J$ into two pieces $J^{(0)}
\in
\H_0$ and $J' \in ({\cal G}\setminus \H_0)=\H'$.  There is a natural
metric
on ${\cal G}$, given by $(A,B)=Str(AB)$ which allows us to make a
canonical
choice of $J'$ such that $Str(J'J^{(0)})=0$.  Now consider the gauge
transformation $g(x) \rightarrow g(x)h(x)$.  Taking into account that
$[\H_0, \H'] \subset \H'$ (the subgroup $H$ is reductive) we get that
$J'$
transforms by conjugation $J' \rightarrow h^{-1}J'h$.  Meanwhile, the
current $J^{(0)}$ transforms inhomogeneously as $J^{(0)} \rightarrow
h^{-1}
J^{(0)} h +h^{-1} dh$.  The inhomogeneous term in this transformation
cancels the inhomogeneous transformation for the gauge field ${\cal
A}$,
such that the action \gauge\ remains invariant.

Integrating out the gauge field ${\cal A}$, we obtain an action for the
sigma model on a coset space
\eqn\act{S_{G/H}^{(0)}={1 \over 4 \pi \lambda^2  } \int d^2 x ~ Str
\big(J'^2
\big)~. }
It is clear that this action is gauge invariant with respect to the
gauge
transformation $g(x) \rightarrow g(x)h(x)$ and therefore is defined on
the
coset space $G/H$.  The group $G$ acts on the coset space by global
left
multiplication, namely for coset representatives $[g]$ and group
element
$g_0 \in G$ we have $g_0 : [g] \rightarrow [g_0 g]$.  It is clear that
$J'$
(as well as $J$) is invariant under this transformation.  Observe that
the
action \act\ can be easily generalized by replacing the $Str()$ by any
ad$(H)$ invariant metric $b(J',J')$. The invariance of the metric
implies
that the action would still possess the same gauge invariance.

The sigma model action \act, however, cannot be the right string
worldsheet
action since this sigma model is not conformal.  At one loop, a counter
term proportional to the Ricci tensor gets generated.  As we will see
in
Section 5, the Ricci tensor for the quotient $PSU(1,1|2)/U(1)\times
U(1)$
is non-zero.

\newsec{WZ term and conformal invariance}

In order to make the theory conformal, we modify the action by adding
an
extra term. This term will be a special version of a WZ term.  Let us
first
remind the reader the structure of the WZ term for a supergroup.  For a
(super) group $G$, the WZ term arises from a closed 3-form.  Indeed,
the
three form reads
\eqn\threeformg{
\tilde \Omega = Str \, (J \wedge  [ \, J \wedge J ]\, ) = 
f_{MNK} \, J^M \wedge J^N \wedge J^K ~,}
where $J^A$ are left-invariant one forms defined as $g^{-1} d g =
\sum_M J^M T_M$, $f_{MNK} \equiv g_{MP} f^P_{NK}$ are the totally
(graded) antisymmetric structure constants, and $g_{MN}$ is a $G$
invariant
bilinear form on ${\cal G}$.  The WZ term is simply the integral of the
pullback of $\tilde \Omega$ over a three manifold whose boundary is the
world sheet.  One can readily verify that $d\,\tilde \Omega =0$ making
use
of the Maurer Cartan identities
\eqn\mc{ d J^K = -{1 \over 2}[J \wedge  J]^K=
-{1\over 2} f^K_{MN} J^M\wedge J^N,}
and the Jacobi identity for ${\cal G}$. The three form should be
closed, so
it is at least locally exact and thus can arise from a two-form $B$.

The possibility of writing the WZ term for the $G/H$ coset arises due
to
the very special nature of the subgroup $H \subset G$. The group $G$
admits
a ${\bf Z}_4$ automorphism and the subgroup $H$ is the fixed locus of
this
action (see for example \con, \zact). This is a general statement which
is
valid for any coset $PSU(n,n|2n)/USp(n,n) \times USp(2n)$.  In this paper,
we
will mainly deal with the $AdS_2 \times S^2$ case ($n=1$) and make some
comments on $n=2$.

Let us denote the projections of the 1-form $J$ on the corresponding
subspaces $\H_k$ of \dec\ as
\eqn\proj{J^{(i)}=J|_{\H_i}~.
}
The naive restriction of the three-form $\widetilde\Omega$ to $G/H$
(obtained by letting indices run only over coset values) is not even
closed\foot{The rules for working with forms and exterior derivatives on 
homogeneous spaces $G/H$ were reviewed in
\ref\doc{E. D'Hoker, {\it Invariant Effective Actions, Cohomology of
Homogeneous Spaces and Anomalies}, Nucl. Phys. B451 (1995) 725,
hep-th/9502162 }.}.
Still, one can write a WZ term using the $\Z$ decomposition of the
algebra.  This WZ
term
is precisely that of \tftwo\ (which is a straightforward generalization
of
the WZ terms in \refs{\ts,\tfol,\tfthree}) and can be written as
\eqn\wzz{\Omega= Str\big(
[J^{(1)}\wedge J^{(1)}] \wedge  J^{(2)} -
[J^{(3)}\wedge J^{(3)}] \wedge  J^{(2)}
\big)~. }
This 3-form is closed and its variation is exact.  Let us introduce the
indices $\{i,j,k, \cdots\}$ for $\H _0$, 
$\{ a,b,c, \cdots \}$ for $\H_1$,
$\{ l, m,n,\cdots \}$ for $\H_2$, and 
$\{ a',b',c', \cdots \} $ for $\H_3$.
Then the 3-form $\Omega$ \wzz\ can be rewritten as
\eqn\theh{
\Omega=  f_{mab} \,J^m \wedge J^a\wedge J^b
-f_{ma'b'}\, J^m \wedge J^{a'}\wedge J^{b'}~.
}
Using the Maurer-Cartan identities \mc\ for the one-forms $J^a$ and $J^{a'}$, 
it is easy to check
that the WZ 3-form $\Omega$ given above is
$d$-exact
\eqn\ex{\Omega= d \,~Str(J^{(1)} \wedge  J^{(3)})  =
d\, (g_{aa'} J^a \wedge J^{a'}) \equiv d\, \Omega^{(2)}.}
The full action can now be written as the sum of the kinetic term
\act, and the (two-dimensional) integral of $\Omega^{(2)}$ 
\eqn\fact{ \eqalign{
S_{G/H} &= S_{G/H}^{(0)} + {i k \over 2 \pi \lambda^2 } \int
\Omega^{(2)} \cr
&= {1 \over 2 \pi \lambda^2 } \int d^2 x
Str \big( \half J_{\mu} ^{(2)} J_{\mu} ^{(2)} +
J_{\mu} ^{(1)} J_{\mu} ^{(3)} +
i k \e ^{\mu \nu}J_{\mu} ^{(1)} J_{\nu} ^{(3)}\big)~.\cr }}
where $k$ will be determined later\foot{The coefficient $i$ in \fact\
makes the contribution of the WZ term to the Euclidean worldsheet
action
real.  This is a little unusual but is forced by conformal invariance
(which will imply that $k$ is real). Moreover, the field $B$ has non
zero
components $B_{a a'}$ and $B_{a' a}$ only in fermionic directions and
we
are not aware of any quantization condition that would require this
term to
be imaginary.}.
Let us stress that this action differs from that of
\refs{\ts{--}\tfthree}\
since it contains a kinetic term for fermions. This kinetic term breaks
$\kappa$-symmetry, however, instead it allows $N=2$ worldsheet
superconformal invariance. To construct the $N=2$ superconformal
generators, it is useful to rewrite the action as (with $\epsilon^{01}
=
+1$)
\eqn\fulact{S_{G/H}={1 \over \pi \lambda^2 } \int d^2 x~
Str \big(  J_{z} ^{(2)} J_{\bar z} ^{(2)} +
(1+k) J_z ^{(1)} J_{\bar z} ^{(3)} +
(1-k) J_{\bar z} ^{(1)}  J_z ^{(3)}\big) .}

Although the coefficient $k$ can be fixed by requiring one-loop
conformal
invariance, we can try to guess the appropriate value of $k$ by
examining
the classical equations of motion.  Taking into account that the
variation
of $J$ satisfies the equation $\delta J=d \delta X+[J,\delta X]$, we
obtain
\eqn\mot{\eqalign{D_{z} J_{\bar z} ^{(3)}+D_{\bar z} J_{z} ^{(3)}
+k \big(\partial_{z} J_{\bar z} ^{(3)}+
             [J_{\bar z} ^{(1)}, J_{z}^{(2)}]+
         [J_{z} ^{(0)}, J_{\bar z}^{(3)}] - (z \leftrightarrow \bar z)
\big)=0\cr
D_{z} J_{\bar z} ^{(1)}+D_{\bar z} J_{z} ^{(1)}
+k \big(\partial_{\bar z} J_{z} ^{(1)}+
             [J_{z} ^{(3)}, J_{\bar z}^{(2)}]+
         [J_{\bar z} ^{(0)}, J_{z}^{(1)}] - (z \leftrightarrow \bar z)
\big)=0\cr
D_{z} J_{\bar z} ^{(2)}+D_{\bar z} J_{z} ^{(2)}
+2 k \big([J_{z} ^{(3)}, J_{\bar z}^{(3)}]-
         [J_{z} ^{(1)}, J_{\bar z}^{(1)}] \big)=0
.}}
For $k=\pm 1/2$, a significant simplification happens.  Using a linear
combinations of the above equations and Maurer-Cartan identities
$\partial_z J_{\bar z}-\partial_{\bar z} J_z +[J_{z}, J_{\bar z}]=0$
and
setting $k=1/2$ we obtain
\eqn\motm{\eqalign{
D_{z} J_{\bar z} ^{(3)}=0 ~~,~~~D_{\bar z} J_{z} ^{(1)}&=0~,\cr
D_{\bar z} J_{z} ^{(2)}-
[J_{\bar z} ^{(1)}, J_{z}^{(1)}] &=0 ~,\cr
D_{z} J_{\bar z} ^{(2)}+
[J_{z} ^{(3)}, J_{\bar z}^{(3)}] &=0~, \cr
}}
where the covariant derivative is defined as $D_{\mu}=\partial_{\mu} +
[J_{\mu}^{(0)},~]$.  So for $k=1/2$, the current $J_z ^{(1)}$ is
covariantly holomorphic and the current $J_{\bar z}^{(3)}$ is
covariantly
antiholomorphic.  (The other choice of $k=-1/2$ would flip the roles of
$J^{(1)}$ and $J^{(3)}$.)  This means that any $H$-invariant
combination of
$J^{(1)}$ currents will be holomorphic when $k=1/2$.

Note that when $k=1/2$, the actions of \fulact\ and \aff\ agree since
$\Pi^c$ is identified with $ J^{(2)}$, ${\widehat\Pi}{}^{\a}$ and 
${\widehat\Pi}{}^{\ad}$ are
identified with $J^{(1)}$, $\Pi^{\a}$ and 
$\Pi^{\ad}$ are identified with $J^{(3)}$.  Furthermore, the four
elements of $J^{(1)}$ decompose under $H=U(1) \times U(1)$ as
$(J^{(1)}_{++}, J^{(1)}_{--}, J^{(1)}_{-+}, J^{(1)}_{+-}$) where the
first
two are identified with $\Pi^\a$ and the second two are identified with
$\Pi^{\ad}$.  So we can define two classically holomorphic, $H$
invariant 
combinations
of
$J^{(1)}$ which are
\eqn\Adef{ A^+ = J^{(1)}_{++}J^{(1)}_{--},\quad 
 A^- = J^{(1)}_{+-}J^{(1)}_{-+}~,}
where we dropped the $z, \bar z$ indices.  We conjecture that the 2d
QFT
theory given by \fulact\ is exactly conformal for $k= 1/2$ and that the
composite fields $A^+(z)$ and $A^-(z)$ remain holomorphic at the
quantum
level. This could be checked by perturbative calculations.
Furthermore, we
conjecture that the OPE's of $A^+$ and $A^-$ form some version of a
$W$-algebra
\eqn\ope{A^+ (z) A^- (w)=-{1 \over (z-w)^4}+{1 \over (z-w)^2}T+
{1 \over (z-w)} W_3 ~,}
where $T$ is the stress-tensor and $W_3$ is some new spin-3 current.
The
OPE's of $A^+$ and $A^-$ with themselves should be regular. The
conformal
anomaly of this $W$-algebra coincides with the superdimension of the
coset
space which is $c=-4$.  
It follows from our calculations of the effective action that the 
W-algebra persist at one-loop level.
Remarkably, it is closely related
to an $N=2$ superconformal algebra and can easily be converted into one
by
adding an extra chiral boson $\rho$.  Let us fix the normalization of
$\rho$ by requiring that $\langle \rho(z) \rho(w) \rangle ={\rm
log}(z-w)$.
Then the fields $e^{\pm i \rho}$ have dimension $\Delta=-1/2$ and the
combinations $G^{\pm}=e^{\pm i \rho} A^{\pm}$ together with
stress-energy
tensor $T+{1 \over 2} \partial \rho\partial \rho$ and $U(1)$ current 
$j=i \partial \rho$ generate a $c=-3$ $N=2$ superconformal algebra.
Adding 
any $N=2$ $c=9$ superconformal field theory for the CY manifold to this
model produces a critical $N=2$ $c=6$ superconformal field theory. This
superconformal field theory is precisely the one described in the
previous
section where $A^+ = (d)^2$ and $A^-=(\bar d)^2$.  It is related to the
RNS
version of the superstring by using the RNS stress tensor, BRST
current,
$b$ ghost, and ghost number current as twisted $N=2$ generators,
redefining
the worldsheet variables, and then untwisting to get a critical $c=6$
superconformal field theory \refs{\mehybrid, \topo}.

\subsec{One loop beta function}

In this subsection we use the background field formalism to compute the
one
loop effective action and verify that there are no UV divergences for
$k=\pm 1/2$.  We write group elements as $g=\tilde g e^{\lambda X}$
where
$\tilde g$ is the background field, $X \in \G$ parameterizes quantum
fluctuations around $\tilde g$, and $\lambda$ is the coupling constant
(i.e.  the inverse radius of $AdS_2 \times S^2$) inserted here for
convenience. Then the current $J_{\mu}$ can be written as
\eqn\curbc{J_{\mu}=g^{-1} \partial_{\mu} g=
e^{-\lambda X} \tilde J_{\mu} e^{\lambda X}+
e^{-\lambda X} \partial_{\mu} e^{\lambda X}~, }
where $\tilde J_\mu={\tilde g}^{-1} \partial_{\mu} \tilde g$ is the
background
current. The action for the coset space is given by \fact\ which in our
parameterization becomes
\eqn\backact{\eqalign{ 
S_{G/H}={1 \over 2 \pi \lambda^2} \int  Str\Big( {1 \over 2} \big(
e^{-\lambda X} \tilde J_{\mu} e^{\lambda X}|_{\G \setminus \H_0} +
e^{-\lambda X} \partial_{\mu} e^{\lambda X} |_{\G \setminus \H_0}
\big)^2+\cr
+ ik \e^{\mu \nu} \big(
e^{-\lambda X} \tilde J_{\mu} e^{\lambda X} +
e^{-\lambda X} \partial_{\mu} e^{\lambda X} \big)|_{\H_1}
\big( e^{-\lambda X} \tilde J_{\nu} e^{\lambda X}+
e^{-\lambda X} \partial_{\nu} e^{\lambda X}\big)|_{\H_3} \Big).
}}

The gauge invariance of the original action $g \rightarrow gh$ (where $h
\in H$) allows us to choose a gauge for $X$ such that $X \in \G \setminus
\H_0$. The gauge invariance of the effective action for the background
field becomes manifest in this gauge for $X$. Under $\tilde g \to \tilde g
h$ we simply change variables $X \to h^{-1} X h$ in the functional
integral. As the subgroup $H$ is reductive, $ h^{-1} X h \in \G
\setminus \H_0$ and so we are still in the same gauge for $X$. The
action \backact\ is invariant under such transformations. The integration
measure is certainly invariant for $k=0$ since there is no chirality.  For
all other $k$, we may use the same regulator as for $k=0$ to see that the
measure is invariant. The gauge invariance of the effective action
guarantees that our theory makes sense on the coset space even
quantum-mechanically.

To compute the effective action for our model, we first have to expand
\backact\ in terms of $X$ and then evaluate all 1PI diagrams with
external lines of the background currents $\tilde J$. To compute the beta
function we need to renormalize UV divergent diagrams. The IR divergences
can be dealt with in the standard fashion by adding a small mass term $\mu$
for $X$. By power counting, the UV primitively divergent diagrams can have
no more than two external lines. Thus we need to evaluate those.

The expansion of \backact\ contains the zeroth order term in $X$ that is
simply the action for the background field and the linear term in $X$ which
does not contribute to 1PI diagrams. We are interested in terms of second
order in $X$ which are the only ones we need to compute the one-loop beta
function.  There is the kinetic term for $X$ that is simply
\eqn\kinq{{1 \over 4 \pi} \int  d^2x~ Str \big(\partial_{\mu} X \big)^2
~,}
and terms which include interactions between $\tilde J$ and $X$.  These
terms can be divided into three subsets: (i) terms containing ${\tilde
J}^{(2)}$, (ii) terms containing ${\tilde J}_z ^{(1)}$ and/or ${\tilde
J}_{\bar z} ^{(3)}$ and (iii) terms containing ${\tilde J}_z ^{(3)}$ and/or
${\tilde J}_{\bar z} ^{(1)}~$\foot{One might also consider terms containing
${\tilde J}^{(0)}$, but the effective action should be independent of them
since their appearance would indicate the breakdown of $H$ gauge
invariance. Using methods similar to those described below, we explicitly
checked that such terms were absent in the one-loop effective action.}.

Terms of type (i) are given by
\eqn\typei{\eqalign{
{ 1 \over \pi} \int d^2x~ Str \Big(&
({1 \over 2}+k) \partial X^{(1)}[{\tilde J}_{\bar z} ^{(2)}, X^{(1)}]+ 
({1 \over 2}-k) \bar \partial X^{(1)}[{\tilde J}_{z} ^{(2)},
X^{(1)}]+\cr
+&({1 \over 2}-k) \partial X^{(3)}[{\tilde J}_{\bar z} ^{(2)},
X^{(3)}]+ 
({1 \over 2}+k) \bar \partial X^{(3)}[{\tilde J}_{z} ^{(2)},
X^{(3)}]+\cr
-&k {\tilde J}_z ^{(2)}[[{\tilde J}_{\bar z} ^{(2)}, X^{(1)}],X^{(3)}]+
k {\tilde J}_z ^{(2)}[[{\tilde J}_{\bar z} ^{(2)},
X^{(3)}],X^{(1)}]+\cr
+&{\tilde J}_z ^{(2)}[[{\tilde J}_{\bar z} ^{(2)}, X^{(2)}],X^{(2)}]
\Big).}}
The first four terms above contain a single background current and
therefore give rise to fish-type divergent diagrams in second order of
perturbation theory. Their combined contribution to the divergent piece
is
equal to
\eqn\fishi{\eqalign{\eqalign{\epsfbox{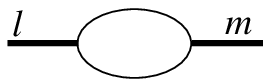} \cr
\noalign{\vskip0.27in } }
\eqalign{=\log(\Lambda/\mu) \,
{\tilde J}_z^l \, {\tilde J}_{\bar z}^m \, 
(&({1 \over 2} - k)^2 ( f_{lab}\, f_{b'a'm}\, g^{bb'} g^{aa'} ) \cr
+&({1 \over 2} + k)^2 ( f_{la'b'}\, f_{bam}\, g^{b'b} g^{a'a} )) ~,}}} 
where $\Lambda$($\mu$) denotes UV(IR) cutoff. Our convention for the
indices and subspaces is the same as defined above equation \theh.  We
put
group-theory factors in parentheses to make comparison easier.

Similarly, the last three terms in \typei\ contain two background lines
and
also renormalize the propagator at one loop. Evaluating their
contribution
to the divergent piece, we obtain
\eqn\propi{\eqalign{\eqalign{\epsfbox{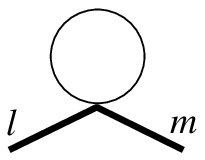} \cr
\noalign{\vskip0.4in }}
\eqalign{=\log(\Lambda/\mu)
{\tilde J}_z^l \, {\tilde J}_{\bar z}^m \, 
\Big( k &( f_{lab}\, f_{b'a'm}\, g^{bb'} g^{aa'} ) \cr
-k &( f_{la'b'}\, f_{bam}\, g^{b'b} g^{a'a} ) \cr
+ &( f_{lik}\, f_{njm} \, g^{kn} g^{ij}) \Big)~,}}} 
The first two terms above have the same group factors as those in
\fishi,
so adding them makes the coefficients in front of both equal to
$1/4+k^2$.
Their sum can be written in terms of the group factor in the third term
of
\propi\ since the groups we choose for our coset constructions have
vanishing dual Coxeter number, which implies, in particular that the
$(l,m)$
components of the Cartan-Killing form vanish:
\eqn\ricci{0 = f_{lab}\, f_{b'a'm}\, g^{bb'} g^{aa'} +
 f_{la'b'}\, f_{bam}\, g^{b'b} g^{a'a} + 2 f_{lik}\, f_{njm} \, g^{kn}
g^{ij} ~.  }
Adding the contributions from both types of diagrams and using \ricci,
we
find that the sum is equal to
\eqn\renormi{({1 \over 2} - 2 k^2 ) \, \log(\Lambda/\mu) \,
{\tilde J}_z^l \, {\tilde J}_{\bar z}^m \, 
( f_{lik}\, f_{njm} \, g^{kn} g^{ij})~.
}
Thus, there is no renormalization only when $k=\pm 1/2$, which are the
same
values of $k$ that we guessed using the classical equations of motion
\mot.

\bigskip 

Repeating the same steps for terms of type (ii) gives
\eqn\typeii{\eqalign{ 
{ 1 \over \pi} \int d^2 x~ Str \Big(
 &({1 \over 2}-{k \over 2}) \partial X^{(2)}
    [{\tilde J}_{\bar z}^{(3)}, X^{(3)}]+ 
  ({1 \over 2}-{3k \over 2}) \partial X^{(3)}
    [{\tilde J}_{\bar z}^{(3)}, X^{(2)}]+\cr
+&({1 \over 2}-{3k \over 2}) \bar \partial X^{(1)}
    [{\tilde J}_{z}^{(1)}, X^{(2)}]+ 
  ({1 \over 2}-{k \over 2}) \bar \partial X^{(2)}
    [{\tilde J}_{z}^{(1)}, X^{(1)}]+\cr
+&(k+1) {\tilde J}_z ^{(1)}[[{\tilde J}_{\bar z} ^{(3)},
X^{(1)}],X^{(3)}]+
   2k\, {\tilde J}_z ^{(1)}[[{\tilde J}_{\bar z} ^{(3)},
X^{(2)}],X^{(2)}]+\cr
+&  k\, {\tilde J}_z ^{(1)}[[{\tilde J}_{\bar z} ^{(3)},
X^{(3)}],X^{(1)}]
\Big)~.}}
The fish-type diagrams give the divergent contribution
\eqn\fishii{\eqalign{\eqalign{\noalign{\vskip0.1in }
\epsfbox{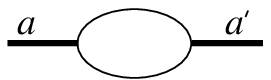}}
\eqalign{={1 \over 2}(1-2k)^2 \, \log(\Lambda/\mu) \, 
{\tilde J}_z^{a} \, {\tilde J}_{\bar z} ^{a'}\, 
(f_{amb} \, f_{b'na'} \, g^{bb'} g^{mn}) ~.}}} 
while the divergent contribution from the last three terms in \typeii\
is
\eqn\fishi{\eqalign{\eqalign{\epsfbox{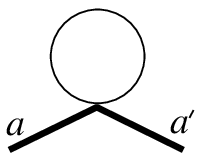}}
\eqalign{ = \log(\Lambda/\mu) \, 
{\tilde J}_z^{a} \, {\tilde J}_{\bar z} ^{a'}\,
\Big(  &3k  (f_{amb} \, f_{b'na'} \, g^{bb'} g^{mn})+ \cr
 &(k+1) (f_{aib'}\, f_{bja'}  \, g^{b'b} g^{ij} ) \Big)  ~.}}} 
Again, the first term above has the same group structure as in \fishii\
while the second can be expressed through it using vanishing of the 
$a, a'$ components of the Cartan-Killing form in a way similar to 
\ricci.  Summing both types of diagrams in this
case gives
\eqn\renormii{(2k^2 - {1 \over 2}) \, \log(\Lambda/\mu) \, 
{\tilde J}_z^{a} \, {\tilde J}_{\bar z} ^{a'}\,
(f_{amb} \, f_{b'na'} \, g^{bb'} g^{mn}) ~
}
which vanishes for the same values of $k=\pm 1/2$.  Terms of type (iii)
in
the expansion of the action can be dealt with in the same manner and
with the same conclusion that $k=\pm 1/2$.

One can immediately deduce from the above results that the one loop
beta
function for $1/\lambda^2$ is proportional to $(2k^2-{1 \over 2})$,
while
the renormalization of $k$ is identically equal to zero independently
of
the values of $\lambda$ and $k$.  We therefore conclude that the {\it
couplings $\lambda$ and $k$ are not renormalized} at one loop. In fact,
we
believe that the the theory given by \fulact\ is exactly conformal for
$k=\pm 1/2$ and we plan to address this issue in the near future.

\newsec{Ricci Curvature, Field Equations and Conformal Invariance}

Here we study the geometrical properties of coset supermanifolds. We
first
show how to compute the Ricci curvature of homogeneous supermanifolds
$G/H$
in terms of structure constants of the associated Lie algebras.  In
doing
this, we will explain the role of the chosen Lie algebra metric. This
will
illuminate the metric interpretation of the work of \bvw\ and \bzv.
Then
we compute the curvature of $G/H$ in terms of the curvature of $G$ plus
additional terms.  These formulas will confirm that
$PSU(1,1|2)/U(1)\times
U(1)$ and $PSU(2,2|4)/SO(4,1) \times SO(5)$ are not flat and therefore
the
corresponding sigma models require some modification for conformal
invariance.

In the second subsection, we present another proof of one-loop
conformal
invariance of our model.  We show that the background fields implied by
the
kinetic term and WZ term satisfy the appropriate target space field
equations that arise from the conditions of one-loop conformal
invariance. This approach makes use of the geometrical properties of
the
coset in question.  We also show that there is no two-loop correction
to
the central charge.

\subsec{Ricci curvature and Lie algebra metrics}

We consider here metrics on homogeneous supermanifolds obtained as
coset
spaces $G/H$ where $G$ is a Lie supergroup and $H$ is a sub(super)group
of
$G$. Our discussion is a generalization of the familiar results for
bosonic
cosets\nref\kn{S. Kobayashi and K. Nomizu, {\it Foundations of
differential
geometry}, vols. I and II. John Wiley \& Sons (1996)}
\nref\hel{S.
Helgason,
{\it Differential Geometry and Symmetric Spaces}, Academic Press, New
York, 1962.}
\refs{\kn, \hel} to the super case. A useful reference is
\ref\dwit{B. DeWitt, {\it Supermanifolds}, Cambridge University Press
(1992).}.

We focus throughout on reductive coset spaces, i.e. spaces where the
Lie
algebra $\G$ can be decomposed as a direct vector space sum of the Lie
algebra $\H$ of $H$ and an ad$(H)$ invariant space $\G \setminus \H$.
We
choose a basis $\{T_M\}$ of generators for $\G$, and use the letters
$\{ M,
N, P,\cdots\}$ to index the generators of $\G$, the letters $\{I, J,
K\ldots\}$ to index the generators of $\calh$ and $\{A,B,C\ldots\}$ for
the
elements of $\G \setminus \H$.

Assume that $\G$ has an ad$(G)$ invariant bilinear form $(T_M, T_N) =
b_{MN}$ whose restriction to $\H$ is non-degenerate and thus can be
used to
produce a reductive orthogonal decomposition $\G=\H \oplus (\G
\setminus
\H)$.  The homogeneous space admits a $G$-invariant metric arising from
the
restriction of the bilinear form $b$ to $\G \setminus \H$.  The
following
result then holds for the curvature of this metric (\kn, vol. 2, p.
203):
\eqn\curr{
R_{ABDC}+ (B \leftrightarrow D) =
{1\over 4} \, f^E_{AB}\, \, b_{EF} \,\, f^F_{CD} + \,
f^I_{AB} \,\, b_{IJ} \,\, f^J_{CD} + (B \leftrightarrow D)\, ~.
}
It is clear that the Riemann curvature tensor does depend on the choice
of
bilinear $b$. To find an expression for the Ricci curvature, we
contract
with the inverse metric $b^{BD}$
\eqn\getric{
R_{AC} = \Bigl(\, {1\over 4} \, f^{E}_{AB}\, \, b_{EF} \,\,f^{F}_{CD}
+ \, f^{I}_{AB} \,\, b_{IJ} \,\, f^{J}_{CD} \Bigr) \, b^{BD}=
  -{1\over 4} \,\, f^{E}_{AD}\,\, \,f^{D}_{BE}
- \, f^{I}_{AD} \,\,\, f^{D}_{CI}
}
The bilinear form drops out of the Ricci tensor as a consequence of its
group invariance and the orthogonality of $\G \setminus \H$ and $\H$.
This
is the desired result for the Ricci curvature of a ordinary homogeneous
manifold $G/H$.  We now claim that for a {\it super coset manifold} we
simply need to replace one of the contractions by a supertrace:
\eqn\supcur{
R_{AB}\, (G/H) =  -{1\over 4} \,\, f^E_{AD}\,\, \,f^{D}_{BE} (-)^E
- \, f^I_{AD} \,\,\, f^{D}_{CI} (-)^I \,.
}
It should be noted that structure constants of a Lie superalgebra are
commuting numbers and thus $f^M_{PQ} =0$ unless $\e(M) = \e (P) + \e(Q)
(\hbox{mod}~2)$.  It follows that $R_{AB}$ vanishes unless both $A$ and
$B$
are commuting or both are anticommuting.  As a consistency check of the
above equation, one can verify the expected exchange symmetry $R_{AB} =
(-)^{AB} R_{BA}$, and that \supcur\ agrees with results derived in
\dwit.
Note that as in the bosonic case, the Ricci curvature is independent of
the
bilinear form.  When $H$ is the identity, the result \supcur\ reduces
to
\eqn\nons{
R_{MN} (G) =
-{1\over 4}\; f^P_{~MQ}\; f^Q_{~NP}\,(-)^P =
-{1\over 4}\, \hbox{Str} \,(T_M T_N) = {1\over 4} \,\, \kappa_{MN} \,,
}
where $T_M$ are the generators of $\G$ in the adjoint representation
${}^P(T_M)_N = f^P_{~MN}$ and $\kappa$ is the Cartan-Killing metric on
$\G$
(positive definite for compact semisimple Lie algebras).  This is an
important fact: {\it the Ricci curvature of a $G\times G$ invariant
metric
on $G$ depends only on the Cartan-Killing form $\kappa$ on $\G$}. This
confirms the metric interpretation of the computation of \refs{\bvw,\bzv}.
Their
$PSU(1,1|2)$ metric arises from the defining representation of
$SU(1,1|2)$ but its Ricci
curvature is still given by the Killing form which vanishes.

\subsec{Further results and the curvature of $PSU(1,1|2)/U(1)\times
U(1)$}

The Ricci curvature of a group $G$, as given in \nons, takes a useful
form
when $G$ has a subgroup $H$.  A small calculation making use of
exchange
symmetries gives
\eqn\red{
R_{AB} (G) =  -{1\over 4}\,f^{M}_{AN}\;f^{N}_{BM}\,(-)^M 
 =  -{1\over 4}\,f^{E}_{AD}\;f^{D}_{BE}\,(-)^E
             -{1\over 2}\,f^{I}_{AD}\;f^{D}_{BI}\,(-)^I\,,
}
where we use the index convention of the previous subsection.  If $H$
defines a symmetric subgroup (i.e. $[\G \setminus \H\, , \, \G
\setminus \H
] \subset \H$), then
\eqn\symred{
R_{AB} (G) =-{1\over 2}\,\,   f^I_{AD} \,\,\, f^D_{BI}\, (-)^I
.}
Combining \supcur\ with \red, we find the following two expressions for
the
Ricci curvature of a coset:
\eqn\firste{
R_{AB} (G/H) = R_{AB} (G)
-{1\over 2}\,\,   f^I_{AD} \,\,\, f^D_{BI}\, (-)^I\,,
}
\eqn\fecfor{
R_{AB} (G/H) = 2R_{AB} (G) +{1\over 4}\,\,   f^E_{AD} \,\,\,
f^D_{BE}\, (-)^E \, .
}
Note that the extra terms involve sums over $\H$ indices in \firste\
and
involve sums over $\G \setminus \H$ indices in \fecfor.

It will also be useful to have an expression for the scalar curvature
$R(G/H)$ of the coset in terms of the scalar curvature $R(G)$ and the
scalar curvature $R(H)$.  Decomposing the scalar curvature of $G$ in
terms
of subgroup and coset indices, we find
\eqn\scalarc{
\eqalign{
R(G) &= R_{IJ} (G) g^{JI} + R_{AB}(G) g^{BA}\,, \cr
&= - {1\over 4} f_{ID}^E \, f^D_{JE} (-)^E g^{JI}
- {1\over 4} f_{IK}^L \, f^K_{JL} (-)^L 
g^{JI}  + R_{AB}(G) g^{BA} \,,\cr
& = - {1\over 4} f_{ID}^E \, f^D_{JE} (-)^E g^{JI} + R(H) +
R_{AB}(G) g^{BA}\,. \cr
}
}
On the other hand, we also have
\eqn\mstpr{ \eqalign{
- {1\over 4} f_{ID}^E \, f^D_{JE} (-)^E g^{JI}
&= - {1\over 4} f_{AD}^I \, f^D_{BI} (-)^I g^{BA}\,,\cr
& = {1\over 2} R(G/H) - {1\over 2} R_{AB}(G) g^{BA} \,,\cr }}
where the first step is proved by raising and lowering indices using
$f_{FID} = g_{FE} f^E_{ID},\,\,f^E_{ID} = (-)^F g^{EF}\,f_{FID}$, and
the
total graded antisymmetry of the structure constants with three lower
indices. The second step made use of \firste. Finally, using \scalarc,
we
find that
\eqn\finscal{
R(G) = {1\over 2}\, R(G/H) +
{1\over 2}\, R_{AB}(G) g^{BA} + R(H) \, .}
\medskip
Let us now consider the computation of the Ricci curvature for the
special
cases.  When $G/H$ is a symmetric coset space, we find from
\fecfor\ that
\eqn\symcons{
R_{AB} (G/H) = 2R_{AB} (G) \qquad  [G/H~ \hbox{symmmetric}]\,.
}
So if we take a Ricci flat supergroup and divide by a symmetric
subgroup,
we get a Ricci flat coset, thus a one-loop conformal sigma model.  The
coset space $G/H = PSU(1,1|2)/ U(1)\times U(1)$, however, is not
symmetric. 
because
$H$ does not contain fermionic generators while there are bosonic
generators in $\G \setminus \H$. Indeed the bracket of a fermionic
generator and any bosonic generator in $\G \setminus \H$ would be a
fermionic generator in $\G \setminus \H$.  Thus, there is no reason to
expect this coset space to be Ricci flat.  To show that the curvature
of
$G/H= PSU(1,1|2)/ U(1)\times U(1)$ is definitely non-vanishing, note
that
$G_B/H$ is a symmetric space, where $G_B= SU(2)\times SU(2)=S^3\times
S^3$
denotes the bosonic subgroup of $G$.  When $G$ is arbitrary, $H$ is
bosonic, and $G_B/H$ is symmetric, \firste\ and \symred\ gives
\eqn\nonvan{
R_{AB} (G/H) = R_{AB} (G)
+ R_{AB} (G_B)\quad (A,B\in \G_B ) .}
So for $PSU(1,1|2)/ U(1)\times U(1)$, we find that
\eqn\finca{
R_{AB} (G/H) =  R_{AB} (G_B) = R_{AB} (S^3 \times S^3) \not= 0\,, }
since the Ricci curvature of round spheres is necessarily nonvanishing.  A
similar result holds for $PSU(2,2|4)/SO(4,1)\times SO(5)$, so it also has
non-vanishing Ricci curvature.

\subsec{Field equations}

We begin our analysis by recalling the familiar background field
equations
that arise from the conformal invariance conditions on two dimensional
bosonic sigma models. When the target space is a manifold with metric
$G_{\mu\nu}$, antisymmetric field $B_{\mu\nu}$ and dilaton $\Phi$,
the sigma model action takes the form \ref\callan{C. Callan, D.
Friedan,
E. Martinec and M. Perry, {\it Strings in background fields},
Nucl. Phys. {\bf B262} (1985) 593;\hfill\break 
E. Fradkin and A. Tseytlin, 
{\it Effective action approach to the superstring theory},
Phys. Lett. {\bf B160} (1985) 69;
\hfill\break A. Sen, {\it Equations of motion for the heterotic string
theory from the conformal invariance of the sigma model}, 
Phys. Rev. Lett. 55 (1985) 1846.}
\eqn\sigmamod{ \eqalign{
S &= {1\over 4\pi \alpha'} \int d\sigma d\tau \sqrt \gamma \gamma^{ab}
\partial_a X^\mu \partial_b X^\nu  G_{\mu\nu} (X) \cr
& \,+ {1\over 4\pi \alpha'} \int d\sigma d\tau \sqrt \gamma
\epsilon^{ab}
\partial_a X^\mu \partial_b X^\nu  B_{\mu\nu} (X) \cr
& \, + {1\over 8\pi}  \int d\sigma d\tau \sqrt \gamma \, R(\gamma) \Phi
(X)\,. }}
The conditions of conformal invariance imply that background fields
satisfy
the equations
\eqn\targeteqn{
\eqalign{
\beta_{\mu\nu}^G &= R_{\mu\nu} - {1\over 4} H^2_{\mu\nu} + 2 \nabla_\mu
\nabla_\nu \Phi= 0~,
\cr
\beta_{\mu\nu}^B  &= {1\over 2} \nabla^\lambda H_{\lambda\mu\nu} -
\nabla^\lambda  \Phi
H_{\lambda\mu\nu} = 0~,
\cr
\beta^\Phi & = {D-26\over 6} + {\alpha'\over 2} \Bigl( - R + {H^2\over
12}
+ 4 (\nabla\Phi)^2 - 4 \nabla^2 \Phi \Bigr) = 0~,
\cr
 }
}
where $\beta^G$ and $\beta^B$ have been calculated to one loop, while
the
$\alpha'$ term in $\beta^\Phi$ arises at two loops.

In the case of the bosonic sigma model, the condition of vanishing beta
function implies the target-space equations of motion. Here we want to
use
this statement when the target space is a supermanifold.  Formally, one
only needs to replace the bosonic indices $\mu,\nu\cdots~$ by indices
$A,
B, \cdots~$ running over bosonic and fermionic values\foot{Sign
factors
that would disappear in the bosonic case are sometimes necessary, and
can
generally be found by considerations of symmetries under exchanges of
indices.}.  Being a part of string theory, the $AdS_n\times S^n$ sigma
model should be supplemented by a set of ghost CFT's and an internal
space
CFT as described in the previous sections.  This cancels the conformal
anomaly, implying that the term $(D-26)$ in $\beta^\Phi$ can be set to
zero. Moreover, for the $AdS_d\times S^d$ model, we expect to have a
constant dilaton so the equations of motion become
\eqn\symplee{
\eqalign{
 \beta^G _{MN}& = R_{MN} + {1\over 4}\; H^P_{~MQ}\; H^Q_{~NP}\,(-)^P =
0\,,
\cr
\beta^B _{NP}&=(-)^M\nabla^M H_{MNP}  = 0\,,
\cr
\beta^{\Phi}& =- {\a' \over 3} R = 0\,.
\cr
 }
}
The first two equations are conditions of one-loop conformal invariance
while the third equation guarantees that the conformal anomaly is not
renormalized at two loops. In fact, these equations are too restrictive
and
can be relaxed without breaking worldsheet conformal invariance.
Namely,
the r.h.s. of these equations can be replaced by the gauge transform of
the
corresponding fields (i.e.  the transformation of the $G$ and $B$ field
under diffeomorphisms plus the gauge transformation of the $B$ field).

Let us begin with the Einstein equation, the first of the above.  For a
WZW
model for a group manifold, comparison with \nons\ shows that setting
$H^M_{~NP} = f^M_{NP}$ makes the equation work. In our case of a coset
manifold $G/H$, taking into account \fecfor\ and $R_{AB}(G)=0$, the
Einstein equation become
\eqn\fecfora{
R_{AB} (G/H) = {1\over 4}\,\,   f^E_{AD} \,\,\,
f^D_{BE}\, (-)^E = -{1\over 4}\; H^E_{~AD}\; H^D_{~BE}\,(-)^E ~.
}
We see that a naive attempt to identify $H_{ABC}$ with structure
constants
would fail.  We can examine our proposal of section 4 to identify the
$H$
field.  The WZ 3-form is given by \theh\ and we will repeat this
expression
here
\eqn\readom{
\eqalign{\Omega  &=H_{ABC} \, J^A\wedge J^B\wedge J^C ~, \cr
&=   f_{mab} J^m \wedge
J^a\wedge J^b -f_{ma'b'} J^m \wedge J^{a'}\wedge J^{b'} ~.
\cr}}
This implies that the only non zero components of $H$ are 
$H_{mab} =  f_{mab}, \, H_{ma'b'} = - f_{ma'b'}\,$, and those with 
the indices $m,a,b$ or $m,a',b'$ being permuted. 
Indeed, the field strength $H$ equals the
structure
constants only up to some crucial sign factors. Raising indices we find
\eqn\hedet{\eqalign{
H_{mb}^{a'} =  f^{a'}_{mb}  \,, \qquad  &H_{mb'}^a= - f_{mb'}^a\,,\cr
 H_{ab}^{m} =  f^{m}_{ab} ~~ \,, \qquad  &H_{a'b'}^m= - f_{a'b'}^m ~ .
}
}
It is easy to check that with the $H$'s listed above \fecfora\ is
satisfied: With indices only running over the coset, and due to the
${\bf
Z}_4$ grading, we can only have $(A,B) = (n,m)$, and $(A,B) = (a,b')$
(or
vice versa).  In both cases the products in the right hand side always
involve an $H$ from the first column in
\hedet\ and an $H$ from the second column. This confirms that the
Einstein
equation \fecfora\ is satisfied.

We now verify the last equation in \symplee\ -- the scalar curvature
$R$ of
the coset supermanifold vanishes. This is readily done with the help of
equation \finscal. Since the numerator group $G$ is Ricci flat, we have
$R(G)=0$ and $R_{AB}(G)=0$. Moreover, the group $H$ is purely bosonic
and
corresponds to $AdS_2\times S^2$ or their higher dimensional
counterparts. While the Ricci curvature of $H$ does not vanish, its
scalar
curvature does. It follows from these facts that $R(G/H)=0$, which is
the
desired result. The central charge of the coset model remains unchanged
to
two loops.

Finally, we turn to the second equation in \symplee; the field equation
for
the three-form.  For the coset case it reads
\eqn\divh{
(-)^A \nabla^A H_{ABC} = g^{AF}\, \nabla_F  H_{ABC} =  g^{AF}\,
(\partial_F
\, H_{ABC} + \cdots )~,}
where the dotted terms denote contributions proportional to the
Christoffel
connection.  Because of homogeneity, it is sufficient to check that
\divh\
holds at the identity point of $G/H$. At this point, to be denoted as
$0$,
the Christoffel coefficients vanish and we have
\eqn\divhc{
(-)^A \nabla^A H_{ABC}\Bigl|_0  =  g^{AF} \,( \partial_F
\, H_{ABC})|_0
.}
It will be useful to treat this left hand side as a two-form,
introducing
the notation
\eqn\calltwo{ {\bf \beta}^{B} \equiv  g^{AF} \,( \partial_F
\, H_{ABC})|_0 \, J^B \wedge J^C \,.
}

We will now show that the $\Z$ automorphism of the group $G$ together
with
its Ricci flatness guarantees that ${\bf \beta}^{B}$ is $d$-exact.
This is
sufficient for one-loop conformal invariance since it implies that the
relevant counter term is a total derivative in the two-dimensional
effective action.  Indeed, as it was shown in section 4, the $\Z$
action
and Ricci flatness are enough to ensure one-loop conformal invariance.
Moreover, we will show that when $H$ is semisimple, ${\bf \beta}^{B}$
vanishes identically. The cases of $AdS_n \times S^n$ for $n=3,5$ are
of
this type. The case $n=2$ is different, but ${\bf \beta}^{B}$ vanishes
anyway.

To evaluate \calltwo, we need the expansion of $H_{ABC}(X)$ to linear
order
in $X$. For this, one uses $J^A = d X^A -\half f^A_{BC} X^B dX^C +
\cdots $,
and, $\Omega = H_{ABC}(X) dX^A\wedge dX^B \wedge dX^C$,  
which lead to
\eqn\hxdep{\eqalign{
H_{ABC} (X) &= H_{ABC}  
 -\half \Bigl(  H_{EBC} \, f^E_{FA} \cr
&\quad + (-)^{AF} H_{AEC} \, f^E_{FB} 
+  (-)^{(A+B)F} H_{ABE} \, f^E_{FC}\Bigr) \, X^F + {\cal O} (X^2)
}}
Replacing into \calltwo\ and noticing that $g^{FA} f_{AFE} =0$, we find
\eqn\newcnd{
 {\bf \beta}^{B} \sim g^{FA} 
\Bigl( (-)^{C(B+F)}\, H_{ACE} f^E_{FB} + (-)^{(B+C)F} H_{ABE} \,
f^E_{CF} 
\Bigr) J^B \wedge J^C \,.
}
For \newcnd\ to be non-vanishing, ${\bf Z}_4$ symmetry implies that
$(B, C)
= (m,n)$, $(B, C) = (b',c)$ or $(B, C) = (b,c')$. For $(B, C) = (m,n)$,
expanding out, substituting the values of $H$, and using the Jacobi
identity, we get an answer proportional to $V_if^i_{mn}J^m\wedge J^n$,
where $V_i \equiv f^a_{ai}$.  For the case $(C, B) = (c, b')$, using
the
Jacobi identity and the vanishing of $R_{cb'} (G)$, we find a single
term
proportional to $V_if^i_{bc'}J^b\wedge J^{c'}$. Evaluating the relative
coefficients of the various terms, one finds
\eqn\botline{
{\bf \beta}^{B} \, \sim \, V_i \,\Bigl( f^i_{mn}\,J^m\wedge J^n
+f^i_{bc'}\,J^b\wedge J^{c'} + 
f^i_{b'c}\, J^{b'}\wedge J^c+  f^i_{jk}\,J^j\wedge J^k \Bigr)~,
}
where the last term has been added by hand using $V_i f^i_{jk} =
f^a_{ai}
f^i_{jk} = 0$, a fact that follows from the Jacobi identity of ${\cal
G}$. Now, using \mc, one can rearrange \botline\ into the $d$-exact
term:
\eqn\divd{
{\bf \beta}^{B}  \, \sim V_i \, d J^i = d \, ( V_i J^i ) \, ,
}
which is sufficient for one-loop conformal invariance since it implies
that
the worldsheet counter term is a total derivative, i.e. it
can be removed by a gauge transformation of the $B$ field.

For the $AdS_2\times S^2$ case, the r.h.s. of \divd\ vanishes since,
using
\spacesplit, one can check that $V_i = f^a_{ai} =0$.  For the higher
dimensional $AdS$ cases the Lie algebra of the subgroup
$H$ is semisimple and therefore every generator $T_i$ can be written as
a commutator
of other generators. As a result, the vanishing of $V_i$ is implied by
the
vanishing of $V_i f^i_{jk}$ discussed above.  This concludes our
verification of the target space equations of motion.

\newsec{String theory in 6 and 10 dimensions}

As we already mentioned in the previous section, our $AdS_2\times S^2$
computations can be repeated for $AdS_3\times S^3$ and $AdS_5\times
S^5$
backgrounds by replacing $PSU(1,1|2)/U(1)\times U(1)$ with $PSU(1,1|2)
\times PSU(2|2)/SU(1,1)\times SU(2)$ or with $PSU(2,2|4)/ SO(4,1)\times
SO(5)$ and choosing the appropriate ${\bf Z}_4$ action. The WZ term is
given by the expression \wzz\ in all of these cases.  Although our
results
only prove one-loop conformal invariance, we conjecture that these
actions
are conformally invariant to all loops and therefore describe conformal
field theories.

The remaining problem for the higher dimensional cases is to construct
a
consistent string theory from the conformal field theory of these coset
spaces. In the six-dimensional case, the conformal anomaly of the coset
model is equal to $c=-10$. As discussed in \refs{\topo,\bvw}, there are
two
ghost-like chiral bosons which are necessary for $N=2$ worldsheet
superconformal invariance and which contribute $c=+2$ to the conformal
anomaly. As will be shown in \me, the remaining $c=+8$ to cancel the
anomaly comes from the ghosts for ``harmonic'' constraints which
eliminate
half of the fermionic currents and reduce the $PSU(1,1|2)\times
PSU(2|2)/SU(2)\times SU(2)$ coset model to the $PSU(2|2)$ model
discussed
in \bvw\ and \bzv.  In the ten-dimensional case, one would have to add
a
$c=+22$ ghost system to cancel the anomaly. Unfortunately, the lack of
a
covariant GS formalism in ten dimensions makes it unclear how to add
this
additional ghost system.  To understand the string theory in $AdS_5
\times
S^5$ background it may be important to analyze the $W$ algebra of the
corresponding coset.

It might be interesting to try to formulate string theory on an
arbitrary
supermanifold. This idea is very simple but is not easy to implement.
Sigma models on supermanifolds are generically non-unitary, so one
needs to
find a set of constraints (analogous to the $N=2$ superconformal
constraints of this paper) to remove the negative-norm states. Although
such constraints are known for some special cases (e.g. those related
to
the RNS superstring by a field redefinition), they are unknown in
general
form.

We have given a formulation of string theory on an interesting
four-dimensional background involving RR fields. We hope our results
will
pave the way to the construction of full string theories for the higher
dimensional cases.

\bigskip
{\bf Acknowledgements:} We would like to thank Dan Freedman, Mark
Grisaru,
Paul Howe,
Vera Serganova, Andy Strominger, Arkady Tseytlin, Cumrun Vafa and 
Edward Witten for useful conversations.
N. B. would like to thank the IAS at Princeton, Harvard University, the
ICTP
at Trieste, and the University of Amsterdam for their hospitality, and
CNPq
grant 300256/94-9 for partial financial support.  
The research of M. B. and S. Z.  was supported by NSF grant
PHY-92-18167 and,
in addition, by the NSF 1994 NYI award 
and the DOE 1994 OJI award.
The work of T. H. and
B. Z. was supported by the US Department of Energy under contract \#
DE-FC02-94ER40818.

\listrefs

\end